\documentclass[12pt]{article}
\pdfoutput=1

\usepackage{amsmath}
\usepackage{amssymb}
\usepackage{graphicx}
\usepackage{cite}
\usepackage{multirow}
\usepackage{longtable}
\usepackage{lscape}

\allowdisplaybreaks[4] 

\newcommand{\capdef}{}
\newcommand{\mycaption}[2][\capdef]{\renewcommand{\capdef}{#2}%
        \caption[#1]{{\footnotesize #2}}}
\makeatletter
\renewcommand{\fnum@table}{\textbf{\tablename~\thetable}}
\renewcommand{\fnum@figure}{\textbf{\figurename~\thefigure}}
\makeatother

\newcommand{\ie}{{\it i.e.}}

\newcommand{\eg}{{\it e.g.}}

\newcommand{\cf}{{\it cf.}}

\newcommand{\eq}{Eq.}

\newcommand{\fig}{Fig.}

\newcommand{\Ref}{Ref.}
\newcommand{\Refs}{Refs.}
\newcommand{\Sec}{Sec.}

\newcommand{\App}{Appendix}

\newcommand{\Tab}{Table}

\newcommand{\equ}[1]{\eq~(\ref{equ:#1})}
\newcommand{\figu}[1]{\fig~\ref{fig:#1}}

\newcommand{\bi}{\begin{itemize}}
\newcommand{\ei}{\end{itemize}}

\newcommand{\be}{\begin{equation}}
\newcommand{\ee}{\end{equation}}
\newcommand{\bea}{\begin{eqnarray}}
\newcommand{\eea}{\end{eqnarray}}

\newcommand{\ldm}{\Delta m_{31}^2}
\newcommand{\sdm}{\Delta m_{21}^2}

\newcommand{\deltacp}{\delta_{\mathrm{CP}}}
\newcommand{\stheta}{\sin^2 2 \theta_{13}}

\begin{document}
\begin{titlepage}

\renewcommand{\thefootnote}{\alph{footnote}}

\vspace*{-3.cm}
\begin{flushright}
EURONU-WP6-10-28  \\
IDS-NF-019 \\
VT-IPNAS 10-19  \\
IFIC/10-50
\end{flushright}


\renewcommand{\thefootnote}{\fnsymbol{footnote}}
\setcounter{footnote}{-1}

{\begin{center}
{\large\bf
Optimization of the Neutrino Factory, revisited\\[0.3cm]
} \end{center}}
\renewcommand{\thefootnote}{\alph{footnote}}

\vspace*{.3cm}
{\begin{center}
{\large \today}
\end{center}
}

{\begin{center} {
                \large{\sc
                 Sanjib~Kumar~Agarwalla\footnote[1]{\makebox[1.cm]{Email:}
                 Sanjib.Agarwalla@ific.uv.es}},
		 \large{\sc
                 Patrick~Huber\footnote[2]{\makebox[1.cm]{Email:}
                 pahuber@vt.edu}},

                 \large{\sc
 		 Jian~Tang\footnote[3]{\makebox[1.cm]{Email:}
                 jtang@physik.uni-wuerzburg.de},
                 Walter~Winter\footnote[4]{\makebox[1.cm]{Email:}
                 winter@physik.uni-wuerzburg.de}}
		 }
\end{center}
}
\vspace*{0cm}
{\it
\begin{center}

\footnotemark[1]
       Instituto de F\'{\i}sica Corpuscular, CSIC-Universitat de Val\`encia, \\
       Apartado de Correos 22085, E-46071 Valencia, Spain \\
\footnotemark[1]${}^,$\footnotemark[2]
       Department of Physics, Virginia Tech, Blacksburg, VA24061, USA \\
\footnotemark[3]${}^,$\footnotemark[4]
       Institut f{\"u}r Theoretische Physik und Astrophysik, 
       Universit{\"a}t W{\"u}rzburg, \\
       D-97074 W{\"u}rzburg, Germany

\end{center}}

\vspace*{0.5cm}

{\Large \bf
\begin{center} Abstract \end{center}  }

We perform the baseline and energy optimization of the Neutrino
Factory including the latest simulation results on the magnetized iron detector
(MIND). We also consider the impact of $\tau$ decays, generated by
$\nu_\mu \rightarrow \nu_\tau$ or $\nu_e \rightarrow \nu_\tau$
appearance, on the mass hierarchy, CP violation, and $\theta_{13}$
discovery reaches, which we find to be negligible for the considered
detector.
For the baseline-energy optimization for small $\stheta$, we
qualitatively recover the results with earlier simulations of the MIND
detector. We find optimal baselines of about $2 \, 500 \, \mathrm{km}$ to $5 \,
000 \, \mathrm{km}$ for the CP violation measurement, where now values
of $E_\mu$ as low as about 12~GeV may be possible. However, for large
$\stheta$, we demonstrate that the lower threshold and the backgrounds
reconstructed at lower energies allow in fact for muon energies as low as
5~GeV at considerably shorter baselines, such as FNAL-Homestake.
This implies that with the latest MIND analysis, low- and high-energy
versions of the Neutrino Factory are just two different versions of
the same experiment optimized for different parts of the parameter
space.
Apart from a green-field study of the updated detector performance, we
discuss specific implementations for the two-baseline Neutrino
Factory, where the considered detector sites are taken to be
currently discussed underground laboratories. We find that reasonable
setups can be found for the Neutrino Factory source in Asia, Europe, and
North America, and that a
triangular-shaped storage ring is possible in all cases based on
geometrical arguments only.

\vspace*{.5cm}

\end{titlepage}

\newpage

\renewcommand{\thefootnote}{\arabic{footnote}}
\setcounter{footnote}{0}

\section{Introduction}

Neutrino oscillation experiments have provided compelling evidence
that the active neutrinos are massive
particles~\cite{GonzalezGarcia:2007ib}, pointing towards physics
beyond the Standard Model. In a three-generation scenario, there are
two characteristic mass squared splittings ($\Delta m_{31}^2\,,\Delta
m_{21}^2$) and three mixing angles
($\theta_{12}\,,\theta_{13}\,,\theta_{23}$) as well as a CP violating
phase $\delta_{\textrm{CP}}$ affecting neutrino oscillations.  A
global fit of solar, atmospheric, accelerator and reactor neutrino
oscillation experiments, yields the following parameter ranges for the
oscillation parameters at the $1\sigma$ level
~\cite{GonzalezGarcia:2010er}: $\Delta
m_{21}^2=(7.59\pm0.20) \times10^{-5}$ eV$^2$, $|\Delta
m_{31}^2|=(2.46\pm0.12) \times10^{-3}$ eV$^2$,
$\theta_{12}=(34.4\pm1.0)^{\circ}$,
$\theta_{23}=(42.8^{+4.7}_{-2.9})^\circ$ and
$\theta_{13}=(5.6^{+3.0}_{-2.7})^\circ$.  Even, within the three
flavor framework, there are still unknowns: the mass hierarchy
(MH)--$\Delta m_{31}^2>0$ (normal ordering) or $\Delta m_{31}^2<0$
(inverted ordering); the value of $\theta_{13}$\footnote{ Note, that
  recent hints for for $\theta_{13}>0$~\cite{Fogli:2008jx} are
  inconclusive and need to await more experimental data.}, and whether
there is CP violation (CPV) in the lepton sector.  The experiment
class, which may ultimately address these questions, is the Neutrino
Factory~\cite{Geer:1998iz,ids}.

In a Neutrino Factory, neutrinos are produced from muon decays in
straight sections of a muon storage ring. The feasibility and a
possible design of a Neutrino Factory have been subject of several,
extensive international studies, such as in
\Refs~\cite{Apollonio:2002en,Albright:2004iw,Bandyopadhyay:2007kx}.
The International Neutrino Factory and Superbeam Scoping
Study~\cite{Bandyopadhyay:2007kx,Abe:2007bi,Berg:2008xx} has laid the
foundations for the currently ongoing Design Study for the Neutrino
Factory (IDS-NF)~\cite{ids}. The goal of the IDS-NF is to present a
conceptual design report, a schedule, a cost estimate, and a risk
assessment for a Neutrino Factory facility by 2013. The IDS-NF defines
a first-version baseline setup of a high energy neutrino factory with
$E_\mu=25 \, \mathrm{GeV}$ and two baselines $L_1 \simeq 3 \,000 - 5
\, 000 \, \mathrm{km}$ and $L_2 \simeq 7 \, 500 \, \mathrm{km}$ (the
``magic'' baseline~\cite{Huber:2003ak}) served by two racetrack-shaped
storage rings, with a muon energy of 25~GeV (for optimization
questions, see, \eg,
\Refs~\cite{Barger:1999fs,Cervera:2000kp,Burguet-Castell:2001ez,Freund:2001ui,Donini:2005db,Huber:2006wb,Gandhi:2006gu,Kopp:2008ds}).
This setup has been demonstrated to have excellent $\stheta$ reaches
for addressing the open questions in the three flavor
scenario~\cite{Huber:2006wb}, to be robust against many potential new
physics effects~\cite{Ribeiro:2007ud,Kopp:2008ds} or systematical
errors~\cite{Tang:2009na}, and to be useful for degeneracy resolution
independently of the finally achieved luminosity~\cite{Huber:2003ak};
for the physics case for the very long baseline, see also
\Ref~\cite{Gandhi:2006gu}.

The appearance signal in a neutrino factory consists of so called
wrong-sign muons (\eg, from $\bar\nu_e \rightarrow \bar\nu_\mu$) and
therefore a detector which is capable of measuring the charge sign of
muons is required in order to distinguish this signal from the
right-sign (\eg, from $\nu_\mu \rightarrow \nu_\mu$) muon background.
The most straightforward solution towards a high fidelity muon charge
measurement is a magnetized iron detector (MIND). With a MIND, the
achievable levels of muon charge identification allow for CP violation
measurements in the muon neutrino appearance
channels~\cite{Cervera:2000kp,Burguet-Castell:2001ez}.

The optimal MIND detector has backgrounds (such as from neutral
currents or charge mis-identification) at the level of about $10^{-3}$
to $10^{-4}$, and the potential to measure the muon charges at
relatively low energies down to a few GeV. The importance of the
precise location of the detection threshold was discussed in detail
in \Refs~\cite{Huber:2002mx,Huber:2006wb}. As the design of the
Neutrino Factory matures, more refined detector simulations have
become available~\cite{Cervera:2010rz,ThesisLaing}, especially in
comparison to the IDS-NF baseline 1.0~\cite{ids}. Compared to the
older analyzes, these new simulations provide the detector response in
terms of migration matrices mapping the incident to the reconstructed
neutrino energy for all individual signal and background channels. An
optimization of the cuts has lead to a lower threshold and higher
signal efficiencies than in previous versions, while the background
level has been maintained in the most recent
analysis~\cite{ThesisLaing}. In addition, separate detector response
functions for neutrinos and anti-neutrinos are available, and it turns
out that the $\bar \nu_\mu$ detection efficiency is better than the
$\nu_\mu$ detection efficiency, which partially compensates for the
different cross sections.\footnote{The difference in neutrino and
  anti-neutrino response is due to the different
  y-distributions~\cite{ThesisLaing}.} The MIND detector has been studied
in \Ref~\cite{Cervera:2010rz,ThesisLaing} as generic neutrino factory
detector and a specific detector of similar type is proposed for the
India-based Neutrino Observatory (INO) to measure atmospheric
neutrinos~\cite{tifr-ino}. The detector at INO may serve as a Neutrino
Factory far detector at a later stage.

Most recently, the background from $\tau$ decays was
discussed for disappearance~\cite{Indumathi:2009hg} and
appearance~\cite{Donini:2010xk} channels. These taus arise from charged
current interaction of $\nu_\tau$ which are due to oscillation, \eg,
for $\mu^+$ stored:
\begin{align}
\text{App.: } &  \nu_e  \rightarrow  \nu_\tau \rightarrow \tau^- \overset{17\%}{\rightarrow} \mu^- \text{ (background) versus }
\nu_e \rightarrow \nu_\mu \rightarrow \mu^- \text{ (signal)} \label{equ:bgtauapp} \\
\text{Disapp.: } &   \bar\nu_\mu  \rightarrow  \bar\nu_\tau \rightarrow \tau^+ \overset{17\%}{\rightarrow} \mu^+ \text{ (background) versus }
\bar\nu_\mu \rightarrow \bar\nu_\mu \rightarrow \mu^+ \text{ (signal)}  \label{equ:bgtaudisapp} 
\end{align}
The reason for these muons to contribute to the background is that
the MIND cannot resolve the second vertex from the $\tau$ decay, in
contrast to OPERA-like emulsion cloud chamber
(ECC)~\cite{Acquafredda:2006ki}. In principle, the muons from $\tau$
decays carry information which may be used for the standard
oscillation~\cite{Donini:2002rm,Autiero:2003fu} or new
physics~\cite{Donini:2008wz} measurements.

An alternative version of the Neutrino Factory with respect to the
IDS-NF baseline has been proposed in
\Refs~\cite{Geer:2007kn,Bross:2007ts,Huber:2007uj,Bross:2009gk,FernandezMartinez:2010zza,Tang:2009wp}.
The key difference is to replace the MIND with a magnetized totally
active scintillator detector (TASD). The TASD, being fully
active, has a lower threshold and better energy resolution.  The
better detector performance and an optimization of the front-end
increasing the intensity have allowed a version of the Neutrino Factory
with $E_\mu \sim5$ GeV and a baseline possibly as short as $L \simeq 1
\, 300 \, \mathrm{km}$, corresponding to FNAL-Homestake. This
version is usually called ``low energy Neutrino Factory'' (LENF) and it is
found that the LENF has especially good performances for large
$\stheta$. In addition, the performance of a TASD allows to exploit
the ``platinum channel'' ($\nu_\mu \rightarrow \nu_e$), however it turns out that it is of little practical value~\cite{FernandezMartinez:2010zza}.
 
The recent simulation results for the MIND have made the performance
margin between TASD and MIND considerably smaller and therefore, we
will show that the distinction between the low- and high-energy
Neutrino Factory is somewhat artificial and merely corresponds to two
extreme corners of a common parameter space.

The phenomenological discussion of the Neutrino Factory has so far
been performed mostly in an abstract baseline-energy space. While the
energy is a continuous variable, it is not obvious that all baselines
can be realized from any accelerator site. Therefore, we will present
a comparison of physics performances for a judicious choice of
accelerator and detector locations. It seems unlikely that a machine
of the size and complexity of the accelerator part of a Neutrino
Factory would be built on a green-field site and therefore, we assume
that it will be co-located with an existing, large accelerator
facility. To be specific we consider: CERN, the Fermi National
Accelerator Laboratory (FNAL), the Japan Proton Accelerator Research
Complex (J-PARC), and the Rutherford Appleton Laboratory (RAL) as
potential accelerator sites~\cite{Berg:2008xx}. For the choice of
potential detector sites the issue is less clear-cut, since, at the
current stage, there is little information on the required amount of
rock overburden for a MIND (or TASD) to perform satisfactorily.
Therefore, we make the conservative choice and assume that a Neutrino
Factory far detector requires a similar amount of rock overburden as
other neutrino experiments do. Under this assumption, a natural choice
of candidate detector sites is given by candidate detector sites for
other neutrino experiments. Fortunately, lists of candidate sites for
general neutrino experiments have been compiled for the US in response
to the National Science Foundation (NSF) call for proposals for a Deep
Underground Science and Engineering Laboratory
(DUSEL)~\cite{DUSELtalk} and for Europe in the context of the Large
Apparatus studying Grand Unification and Neutrino Astrophysics
(LAGUNA)~\cite{laguna} study. In North America, we consider eight
locations: Soudan, WIPP, Homestake, SNOLAB, Henderson, Icicle Creek,
San Jacinto, and Kimballton. In Europe, under LAGUNA, there are seven
possible candidate sites: Pyh{\"a}salmi in Finland, Slanic in Romania,
Boulby in UK, Canfranc in Spain, Fr{\'e}jus in France, SUNLAB in
Poland, and Umbria in Italy. Along with these seven sites, we also
consider Gran Sasso National Laboratory (LNGS) in Italy and Gran
Canaria in Spain.  We will complement these lists of detector sites by
the Asian facilities: the Kamioka mine in Japan, the proposed Chinese
underground laboratory at CJPL, YangYang in Korea, as well as INO in
India.

This paper is organized as follows: We describe our
methods and implementation in \Sec~\ref{sec:setup-comparison}. After
that, we update the simultaneous optimization of baseline and muon
energy in \Sec~\ref{sec:LvsE} in a green-field scenario. In
\Sec~\ref{sec:site}, we discuss the selection of specific sites, the
site geometry, and the possibility to use a triangular-shaped storage
ring. Furthermore, in \Sec~\ref{sec:siteperf}, we quantify
site-specific performance of the Neutrino Factory. Finally, we
summarize and draw our conclusions in \Sec~\ref{sec:conclusion}.  
Details for the assumptions for the individual accelerator and detector sites
can be found in \App~\ref{app:sites}. The
sensitivity curves for all possible considered site combinations are
given \App~\ref{app:baseline-combinations}. The individual data files for the curves
are available for download at \Ref~\cite{dpage}.

\section{Simulation method and performance}
\label{sec:setup-comparison}

In this section we describe our simulation method and we show the
difference to the IDS-NF 1.0 in terms of event rates. We also
compare the performance resulting from the different detector
simulations, and we compare the performance between one and two
baselines.

\subsection{Simulation method}

For the simulation of the Neutrino Factory, we use the GLoBES
software~\cite{Huber:2004ka,Huber:2007ji}.  The description of the
experiment is based on \Refs~\cite{Huber:2002mx,Huber:2006wb}, where
we use the parameters from the IDS-NF baseline setup 1.0 (IDS-NF 1.0)
described in \Ref~\cite{ids} (note number IDS-NF-002). The detector description of
this setup is based on \Ref~\cite{Abe:2007bi}, which has been updated
in \Refs~\cite{Cervera:2010rz,ThesisLaing}. In this section, we
compare these three detector descriptions, whereas we use only the
most recent version, \Ref~\cite{ThesisLaing}, in the following
sections. IDS-NF 1.0 uses two magnetized iron calorimeters (fiducial
mass 50~kt) at $L= 4000~{\rm km}$ and $L= 7500~{\rm km}$. There are
two racetrack-shaped storage rings pointing towards these detectors,
with a luminosity of $2.5 \times10^{20}$ useful muon decays per
polarity, decay straight, and year, \ie, $10^{21}$ useful muon decay
per year. We assume a running time of 10~years, \ie, $10^{22}$ useful
muon decay in total.  The parent muon energy is assumed to be $E_\mu
=25 \,\mathrm{GeV}$.  The considered oscillation channels are:
\begin{align}
\nu_\mu \text{ appearance:   } & \nu_e \rightarrow \nu_\mu \text{ for } \mu^+ \text{ stored} \, , \\
\bar\nu_\mu \text{ appearance:   } & \bar\nu_e \rightarrow \bar\nu_\mu \text{ for } \mu^- \text{ stored} \, ,\\
\nu_\mu \text{ disappearance:   } & \nu_\mu \rightarrow \nu_\mu \text{ for } \mu^- \text{ stored} \, ,\\
\bar\nu_\mu \text{ disappearance:   } & \bar\nu_\mu \rightarrow \bar\nu_\mu \text{ for } \mu^+ \text{ stored} \, . 
\end{align}

Since the luminosity changes if one or two storage rings are required,
\ie, one or two baselines are operated, and the efficiency of a
triangular-shaped ring, which will discuss later, is different, it is
convenient to re-parameterize luminosity in terms of a scale factor
(SF)~\cite{Tang:2009wp}: SF=1 corresponds to the above mentioned
parameters $2.5 \times10^{20}$ useful muon decays per polarity, decay
straight, and year. If only one baseline is needed, then all muons can
be injected in the same storage ring, and SF=2. If, on the other hand,
a storage ring with a different geometry (such as a triangle) is used
to point towards the two baselines simultaneously, all muons will be
injected into this ring, but the straight length towards each detector
will be smaller than in the racetrack case, \ie, 0$<$SF$<$2 in
general. The scale factor is then convenient to parameterize the
obtained luminosity relative to the IDS-NF baseline setup: SF$>$1:
higher luminosity, SF$<$1: lower luminosity. Note that, in principle,
the SF can, for lower $E_\mu$, also be increased by a re-optimization
of the front-end and generally will increase for lower energies due to
the reduced decay losses during acceleration. For example, a SF=2.8
for a low energy 4~GeV Neutrino Factory has been obtained in
\Ref~\cite{Bross:2009gk} compared to SF=2.0. We will not consider this
type of effect, since it depends on the accelerator complex in a
non-trivial fashion.

For the updated detector simulations, we use the migration matrices
mapping the incident to the reconstructed neutrino energies for all
individual signal and background channels, which can be directly
implemented into GLoBES. Note that charge mis-identification,
(electron) flavor mis-identification and neutral current backgrounds
are included. For the binning, we then follow
\Ref~\cite{Cervera:2010rz,ThesisLaing}, where the migration matrices
for the appearance channels are given. For the disappearance channels,
we use the same matrices.\footnote{That is somewhat on the conservative
  side, since we require charge identification and better results may
  be obtained with an event sample without charge
  identification~\cite{Huber:2006wb}.} In addition, we increase the
number of sampling points for high energies to avoid aliasing. This
implementation will be used throughout the remainder of this paper,
unless indicated otherwise. It is denoted by the label
``new-NF''. Note that we also include signal (2.5\%) and background
(20\%) normalization errors, uncorrelated among all oscillation channels.

For the $\nu_\tau$ contamination, we use the migration matrix from
\Ref~\cite{Donini:2010xk} for both the $\nu_e \rightarrow \nu_\tau$
and $\nu_\mu \rightarrow \nu_\tau$ channels, since it only depends on
characteristics of the $\tau$ decays. Note that, since the binning
given in there is different from
\Refs~\cite{Cervera:2010rz,ThesisLaing}, we had to re-bin this matrix
carefully. As an important consequence, all events below 2~GeV are
collected in the lowest bin.  We also apply the muon kinematic cuts
for the muons from the $\tau$ decays as for the golden channel, 
following \Ref~\cite{Donini:2010xk}. In a
more refined approach, one may want to have the migration matrices
from incident $\nu_\tau$ energy to reconstructed $\nu_\mu$ energy
directly. This setup will be denoted as ``new-NF$\tau$'' and it
contains everything in new-NF plus the muons from $\tau$ decays.  As
we will show new-NF$\tau$ produces practically the same results as
new-NF\footnote{This statement is true only for the performance
  indicators used in this paper, which all focus on the the appearance
  channel, and will most likely not apply to precision measurements
  of the atmospheric neutrino parameters in the disappearance
  channels as indicated in \Ref~\cite{Indumathi:2009hg}.}.

The input oscillation parameters are taken as
follows~\cite{GonzalezGarcia:2010er}, unless noted otherwise:
\begin{align}
  &\theta_{12}=34.4^\circ\,,\quad\theta_{13}=5.6^\circ\,,\quad\theta_{23}=42.8^\circ
  \nonumber\\
  &\Delta m_{21}^2=7.59\times10^{-5} \, \text{eV}^2\,,\quad |\Delta m_{31}^2|=2.46\times10^{-3} \, \text{eV}^2\,.
\label{equ:params}
\end{align}
We impose external $1\sigma$ errors on $\Delta m^2_{21}$ (4\%) and
$\theta_{12}$ (4\%) and on $\Delta m^2_{31}$ (10\%) and $\theta_{23}$
(10\%) as conservative estimates\footnote{Here we expect that 
the best measurement of the atmospheric parameters comes from 
the Neutrino Factory in the near future. However,it is useful to add the current information on the atmospheric parameters to speed up the marginalization and degeneracy finding.} 
for the current measurement errors~\cite{GonzalezGarcia:2010er}. We also include a 2\% matter
density uncertainty~\cite{Geller:2001ix,Ohlsson:2003ip}. Unless noted
otherwise, we simulate the normal hierarchy.

\subsection{Event rate comparison}
\label{subsec:spectra}

\begin{figure}[tp]
 \centering
 \includegraphics[width=\textwidth]{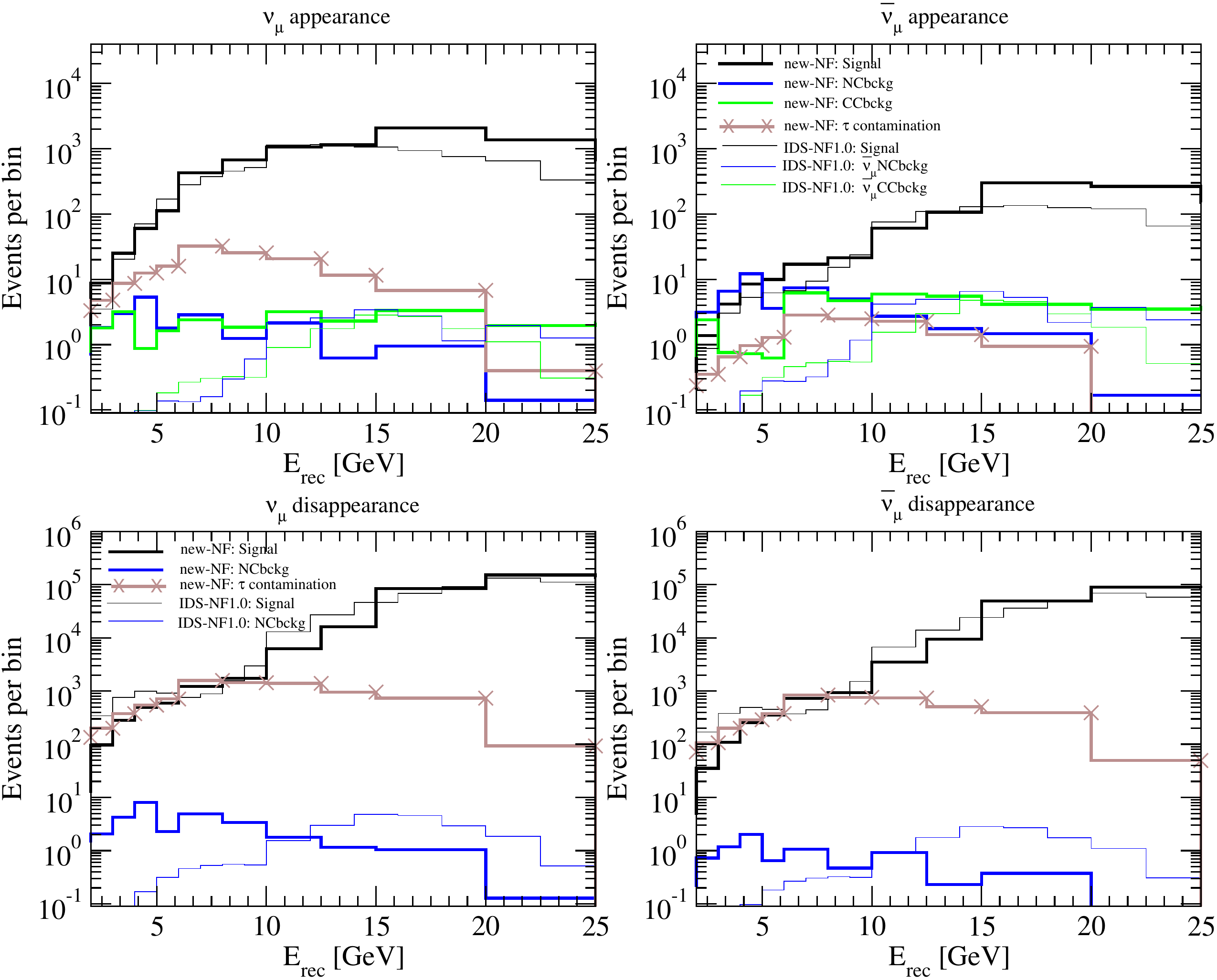}
 \mycaption{A comparison of the event rate spectra between
   new-NF~\cite{ThesisLaing}  (thick
   curves), including backgrounds
   from $\nu_\tau$~\cite{Indumathi:2009hg,Donini:2010xk},
    and IDS-NF 1.0 (thin curves) for the different
   oscillation channels as given in the plot legend. The chosen
   oscillation parameters are taken from \equ{params} with
   $\deltacp=0$. The muon energy is $25$ GeV and the detector mass is
   50kt at a baseline of $4000$~km.}
 \label{fig:spectra}
\end{figure}
\begin{table}[t!]
 \begin{center}
 \begin{tabular}{|l|r|rrr|}
   \hline
   & Signal & NC bckg & CC bckg & $\nu_\tau$ bckg\\
   \hline
   $\nu_\mu$ (app)&7521 & 20 & 25 & 142 \\
   $\bar{\nu}_\mu$ (app) & 924 & 45 & 39 & 13 \\
   $\nu_\mu$ (disapp) & $4.0\times10^5$ & 31 & - &8154  \\
   $\bar{\nu}_\mu$ (disapp)&$2.4\times10^5$ & 8 & - & 4337  \\ 
   \hline
\end{tabular}
\end{center}
\mycaption{\label{tab:events} The expected event rates for new-NF$\tau$ in a 50kt detector  at a 4000~km baseline with a muon energy of $25$ GeV. The chosen oscillation parameters are taken from \equ{params} with $\theta_{13}=5.6^\circ$ and $\deltacp=0$. }
\end{table}
In \figu{spectra}, we compare the event rates of the latest detector
simulation new-NF (thick solid curves) with IDS-NF
1.0 (thin solid curves) for the four different oscillation channels as
given in the plot legend. 

IDS-NF 1.0 (thin curves) did not use any migration matrices and this
is reflected in the background shape, both neutral current (NC) and 
charged current (CC), which closely
follows the signal shape. The signal shape of IDS-NF 1.0 is quite
similar to the one of new-NF, indicating that migrations are not large
for the signal, which is not surprising since energy reconstruction
works well for the signal events. The background shapes, on the other
hand, differ substantially between IDS-NF 1.0 and new-NF, since here
migrations are non-negligible. In particular for the NC background, we
observe that for new-NF (thick curves) it is quite peaked at low
energies. This phenomenon is known as ``feed-down'': for a given
incoming neutrino energy, there will be less energy deposited in the
detector in a NC event than in a CC event, simply because a neutrino
is leaving the detector carrying away a sizable fraction of the
incoming energy. If a NC event is mis-identified as being a CC
event\footnote{Otherwise, it would not be a background event.}, then
the CC event kinematics will be used for energy reconstruction, which
assumes that
$E_\nu^\text{rec}=E_\text{lepton}^\text{rec}+E_\text{hadrons}^\text{rec}$.
This results in a systematic downward bias in the reconstructed energy for
NC background events. This feed-down is the strongest effect of
migration and thus has potential impact on the energy optimization,
since it penalizes neutrino flux at high energies, where there is
little oscillation but a large increase in fed-down background.
Also, for muons from $\tau$ decays there is a strong feed-down for a
similar reason: in the decay of a $\tau$ there will be two additional
neutrinos which leave the detector. Here, the disruptive effect of
high energies is even more pronounced, since the $\nu_\tau$ CC cross
section is a steeply increasing function of neutrino energy up to
about 30~GeV.

In summary, the CC backgrounds in new-NF pile-up at lower energies.
These low energy events are relevant for degeneracy resolution,
especially for intermediate values of $\stheta \sim 10^{-4} -
10^{-2}$. However, the oscillation peak in vacuum would be at about
10~GeV, and matter effects are most important at about 8~GeV, which
need to be covered especially for small $\stheta$, where the event
rates otherwise rapidly decrease with distance. The backgrounds from
$\tau$ decay in new-NF$\tau$ tend to collect around 8~GeV and may
present an immediate problem for all values of $\stheta$. Therefore,
it is not quite clear that high muon energies are preferred
everywhere in the parameter space, and one may suspect that the
baseline-muon energy optimization may be a complicated function of the
detector response.

\subsection{Performance and impact of $\boldsymbol{\nu_\tau}$ contamination}
\label{subsec:performance}

Neutral current backgrounds do not carry any information about flavor
conversions of active neutrinos and therefore are detrimental to
oscillation searches. The muons from $\tau$ decays, on the other hand,
do arise from oscillation and they are a sign of appearance of a new
flavor, $\tau$, in a beam otherwise devoid of this flavor. The
background arising from $\nu_\tau$ as defined in \equ{bgtauapp}
(appearance channels) and \equ{bgtaudisapp} (disappearance) are shown
as gray (brown) solid curves in \figu{spectra}. In all channels, they
are the largest source of background. It is, however, not clear from
the beginning whether this is a benefit or a curse, since this
oscillating background carries information on the oscillation
parameters. In particular, the low energy parts, which actually stem
from much higher incident neutrino energies, may carry complementary
information to the high energy signal; since the resulting energy
distribution is different they may be separated on a statistical
basis. For example, the $\nu_\mu$ appearance probability is given,
expanded to second order in $\sin 2 \theta_{13}$ and the hierarchy
parameter $\alpha \equiv \sdm/\ldm \simeq 0.03$,
as~\cite{Cervera:2000kp,Freund:2001pn,Akhmedov:2004ny}:
\begin{eqnarray}
 P_{e\mu} &\simeq& 
 \sin^22\theta_{13} \sin^2\theta_{23} 
\frac{\sin^2[(1-\hat{A})\Delta_{31}]}{(1-\hat{A})^2}\nonumber \\
&\pm& \alpha \sin2\theta_{13} \sin\deltacp \sin2\theta_{12} \sin2\theta_{23}  \sin(\Delta_{31}) \frac{\sin(\hat{A}\Delta_{31})}{\hat{A}}
\frac{\sin[(1-\hat{A})\Delta_{31}]}{(1-\hat{A})} \nonumber \\
&+& \alpha \sin2\theta_{13} \cos\deltacp \sin2\theta_{12} \sin2\theta_{23}  \cos(\Delta_{31}) \frac{\sin(\hat{A}\Delta_{31})}{\hat{A}}
\frac{\sin[(1-\hat{A})\Delta_{31}]}{(1-\hat{A})} \nonumber \\
&+& \alpha^2 \cos^2\theta_{23} \sin^22\theta_{12} 
\frac{\sin^2(\hat{A}\Delta_{31})}{{\hat{A}}^2} 
\label{equ:papp}
\end{eqnarray}
with $\Delta_{31} \equiv \ldm L/(4E)$ and $\hat{A} = \pm 2 \, \sqrt{2}
\, E \, G_F \, n_e/\Delta m_{31}^2$. The signs in the second term and
$\hat{A}$ are positive for neutrinos and negative for anti-neutrinos.
For $P_{e \tau}$, the channel which controls the background in
\equ{bgtauapp}, flip the sign of the second and third terms and
replace in the first and fourth terms $\sin^2 \theta_{23}
\leftrightarrow \cos^2 \theta_{23}$. For maximal atmospheric mixing,
only the signs of the second and third terms change. Now consider, for
instance, the magic baseline $L \simeq 7 \, 500 \, \mathrm{km}$ where,
by definition, $\sin ( \hat A \Delta_{31} ) \simeq
0$~\cite{Huber:2003ak}. In this case, only the first term survives,
which is the same for the signal and for the background, which means
that it adds to the $\stheta$ and MH sensitivity. For the short
baseline used for the CPV measurement, the sign of the second and
third terms are different between $P_{e \mu}$ and $P_{e \tau}$, which
means that the effects of $\deltacp$ are, naively, reduced by the
$\nu_\tau$ background. However, note that the background is
reconstructed at lower energies, which means that one can, in
principle, distinguish the two channels. It is therefore, without
numerical simulation, not obvious if the $\nu_\tau$ contaminations
improve or deteriorate the sensitivities. 
\begin{figure}[tp]
 \centering
 \includegraphics[width=\textwidth]{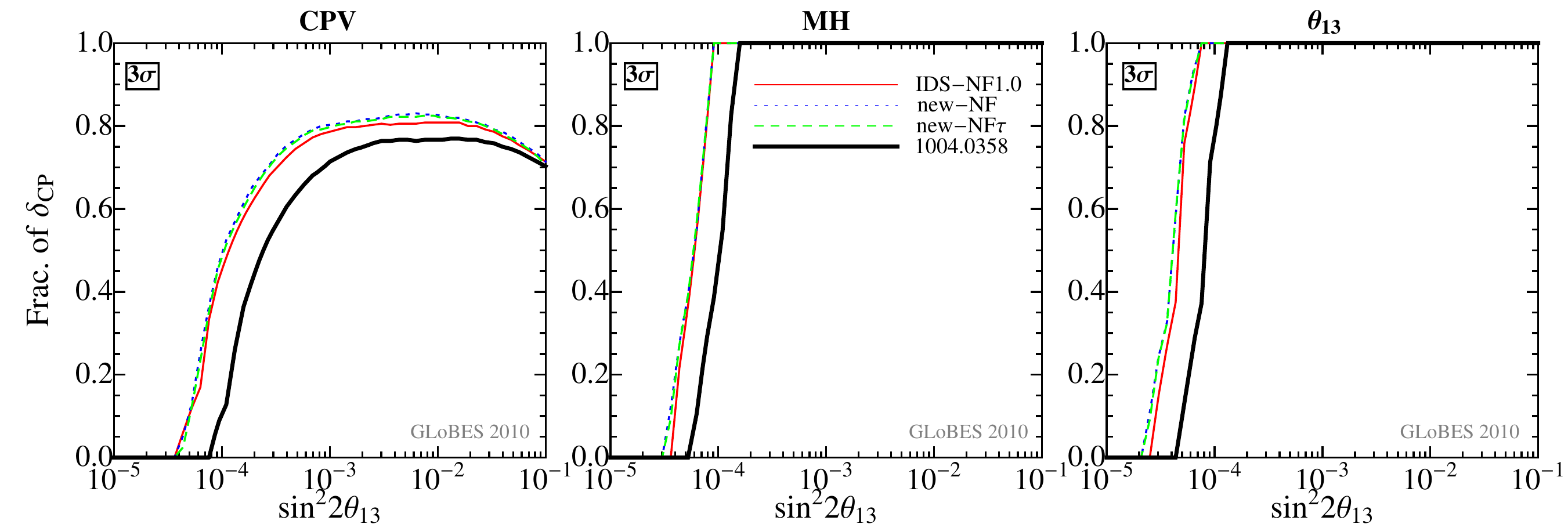}
 \mycaption{A comparison of the discovery reach of CPV, MH, and
   $\theta_{13}$ at the $3\sigma$ CL among different detector
   simulations. The label ``IDS-NF'' refers to the detector in the
   IDS-NF baseline setup 1.0~\cite{ids}. The simulation with the
   migration matrices from~\cite{Cervera:2010rz} is indicated by the
   label ``1004.0358''. The label ``new-NF'' refers to most up-to-date
   detector simulation in \Ref~\cite{ThesisLaing}. The $\nu_\tau$
   contaminations in the appearance and disappearance channels are, in
   addition, included in ``new-NF$\tau$''~\cite{Indumathi:2009hg,Donini:2010xk}. Here
   a combination of two baselines $4\, 000$ km and $7 \, 500$ km with
   two 50~kt MIND detectors is assumed.}
 \label{fig:performance}
\end{figure}

The physics performance arising from the different detector
simulations for the CPV, MH, and $\theta_{13}$ discovery reaches are
show in \figu{performance}. Here ``IDS-NF 1.0'' refers to the detector
performance of the IDS-NF baseline setup 1.0~\cite{ids}. The results
in the figure demonstrate that the performance based on the detector
simulation presented in \Ref~\cite{Cervera:2010rz} (thick solid curves)
is worse.  The main reason, we were able to identify, is significantly
higher backgrounds from charge mis-identification than in the IDS-NF
1.0. The most up-to-date
detector simulation is presented in \Ref~\cite{ThesisLaing} (dotted
curves) and this setup is labeled new-NF, for which the performance is
slightly better than for the IDS-NF 1.0.  In this case, the signal
efficiencies and threshold are improved compared to IDS-NF 1.0, while
the background level is maintained. One of the main differences with
respect to \Ref~\cite{Cervera:2010rz} is the inclusion of
quasi-elastic events which improves the signal efficiency at low
energies. The effect of the migration of the backgrounds does not have
a large impact on the discovery reaches. This may not be true for
precision studies of the atmospheric oscillation parameters, however,
a detailed answer to this question is beyond the scope of the current
paper. If, in addition, the contributions from the $\nu_\tau$ are
included, new-NF$\tau$ (dashed curves), there is hardly any effect on
the performances.  Note, that the relative impact of $\tau$ decays
does depend on the underlying detector parameters and this illustrates
that it is difficult to predict the effect of the $\nu_\tau$ without
numerical simulation. In any case, the absence of a significant
difference in performances between new-NF and new-NF$\tau$ is in
agreement with the results presented in \Ref~\cite{Donini:2010xk} and
therefore, we will not further consider $\tau$ decays and the
resulting backgrounds.

\begin{figure}[tp]
 \centering
 \includegraphics[width=\textwidth]{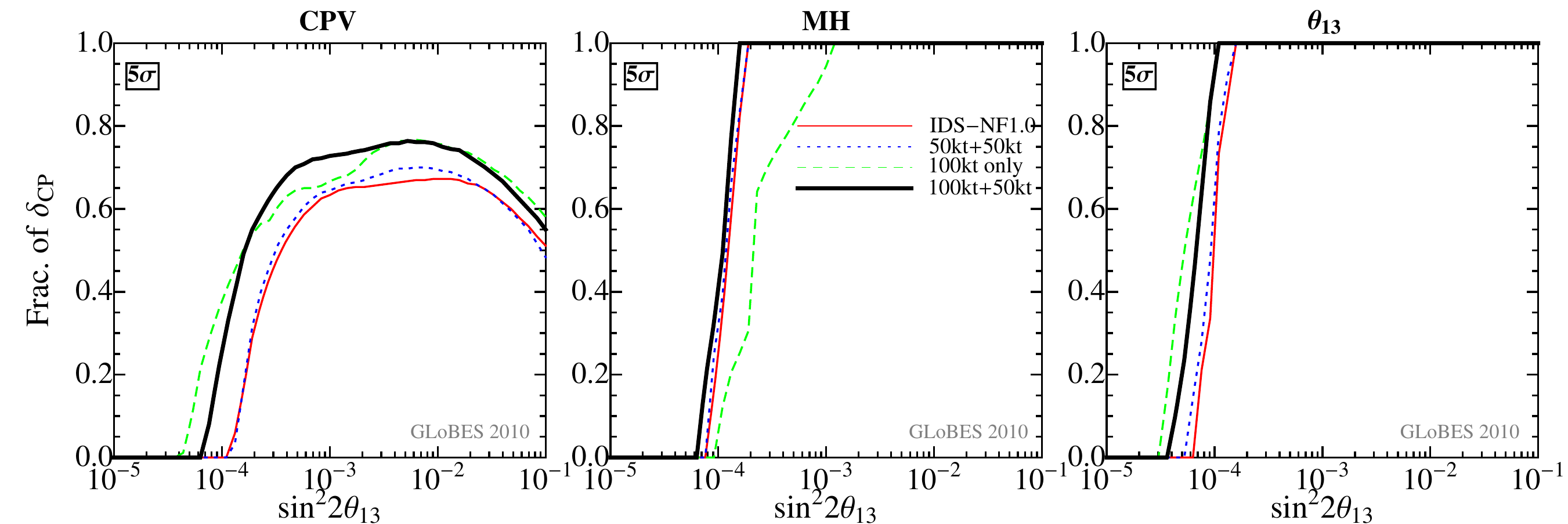}
 \mycaption{\label{fig:detector} A comparison of the discovery reach
   of CPV, MH, and $\theta_{13}$ at the $5\sigma$ CL among different
   experimental setups: ``50kt+50kt'' refers to a combination of 50~kt
   MIND at $4 \, 000$ km and 50~kt MIND at $7 \, 500$ km (SF=1),
   ``100kt only'' to a 100~kt MIND at $4 \, 000$ km (SF=2),
   ``100kt+50kt'' to the combination of a 100~kt MIND at $4 \, 000$ km
   and 50~kt MIND at $7 \, 500$ km (SF=1). All these setups use the
   most up-to-date detector simulation new-NF~\cite{ThesisLaing}.
   ``IDS-NF 1.0'' refers to IDS-NF baseline setup~1.0, \ie, the
   combination of 50~kt MIND at $4 \, 000$ km and 50~kt MIND at $7 \,
   500$ km (SF=1), using no migration matrices~\cite{ids}  (note number IDS-NF-002),
to be compared to the dotted curves. }
\end{figure}

Other questions to be addressed in the context of the updated detector
simulation are the quantitative comparison between one and two
baselines, and the impact of a larger detector at the shorter
baseline. We discuss these in \figu{detector}, where several versions
of the updated detector are compared with the IDS-NF 1.0.
Note that the scale factor (SF) has been adjusted for the assumed
racetrack storage rings to correct for the larger number of useful
muon decays during the single baseline operation. In addition, note
that this figure is shown at the $5 \sigma$ CL, compared to the
previous, to make the impact of degeneracies clearer.
From the comparison of the IDS-NF 1.0 and the corresponding 50kt+50kt
curves using new-NF confirm the earlier result, note that there is not much
difference in performance.  A possible alternative setup is to operate
a single 100~kt detector at the $4 \, 000 \, \mathrm{km}$ baseline,
this configuration is labeled ``100kt only''. This configuration
actually exhibits better performances for CPV and the $\theta_{13}$
discovery because of the factor of two higher luminosity using the
racetrack-shaped storage rings. In this case, the complementary
information at the $7 \, 500 \, \mathrm{km}$ baseline is replaced by
high statistics at the short baseline. Note, however, that the MH
discovery reach is significantly worse, and that degeneracies affect
the shape of the CPV curve. The setup ``100kt+50kt'', where there is a
100~kt detector at 4000~km and one 50kt at 7500~km, can easily resolve
the degeneracies at about $\stheta \sim 10^{-3}$ in the CPV discovery
reach, while the MH and $\theta_{13}$ discovery reaches are
comparable.  In this case, SF=1, which means that this setup in fact
only has 75\% of the exposure of the ``100kt only'' version.
Therefore, the two baselines are synergistic in the sense of
\Ref~\cite{Huber:2002rs}, \ie, for the same exposure, the baseline
combination clearly performs better.  However, if one sticks to the
racetrack geometry of the storage rings, the one baseline operation
may be more efficient. For a triangular shaped ring, which we will
discuss later, this argument changes, because the second baseline is
available anyway. The question of the necessity of the magic baseline
remains open. Especially in the context of new physics and surprises,
such as a lower than expected machine luminosity, it provides a robust
alternative.

In the following, now that we have quantified the impact on the
performance, we will only consider the setup with the updated
migration matrices from \Ref~\cite{ThesisLaing}, \ie, new-NF.
Wherever we refer to ``IDS-NF'', we will actually mean the IDS-NF parameters
($E_\mu=25 \, \mathrm{GeV}$, $4 \, 000 \, \mathrm{km}+7 \, 500 \,
\mathrm{km}$), while the detector simulation is new-NF. We will not consider
the $\nu_\tau$ contribution anymore, partially because it has been
shown not to have a significant impact on the discussed performance
indicators, partially because it will need to be quantified within the
same detector simulation as the signal and other backgrounds in the
future.

\section{Optimization of a green-field setup, low versus high energy
  Neutrino Factory?}
\label{sec:LvsE}

Here we study the optimization of a green-field setup, which means
that no particular accelerator and detector sites are chosen and 
that the baselines and muon energy are not
constrained. The optimization is performed using the migration
matrices from \Ref~\cite{ThesisLaing}. Now that the detection
threshold has improved, we are especially interested if the new MIND
detector can interpolate between low and high energy Neutrino Factory.

\begin{figure}[tp]
 \centering
 \includegraphics[width=\textwidth]{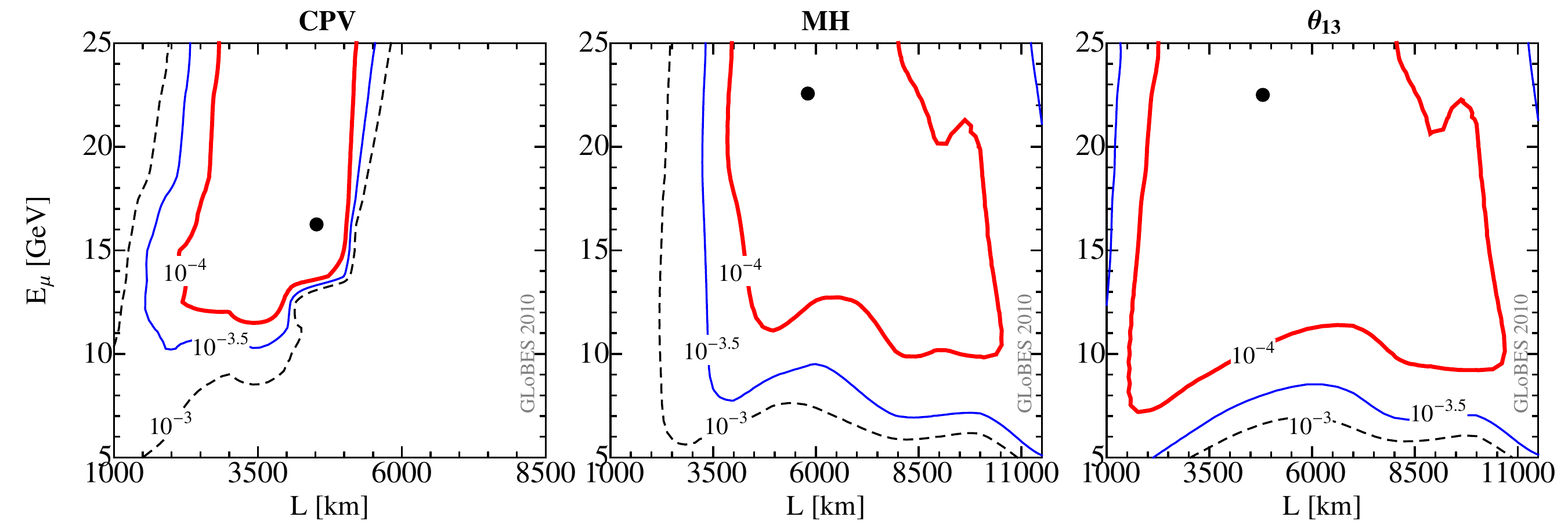} 
 \mycaption{\label{fig:LvsE-th13} Discovery reach in $\stheta$ for
   maximal CP violation, MH, and $\theta_{13}$ as a function of
   baseline and $E_\mu$. The contours show for how small (true)
   $\stheta$ the different quantities will be discovered at the $3
   \sigma$ CL, where maximal CP violation $\deltacp=\pi/2$ is chosen
   as a true value in all cases. The best reaches for baseline and $E_\mu$ 
   are marked by dots: (4519,16.25), (5805,22.57) and (4800,22.50) followed 
   by their optimal sensitivity of $\stheta$ at $10^{-4.8}$, $10^{-4.5}$ and $10^{-4.5}$. 
   Here SF=1 is used with one 50 kt detector. }
\end{figure}

First of all, consider that $\stheta$ is not found before the Neutrino Factory operation. Assume that, in this case, one wanted to optimize for the reach in $\stheta$, \ie, CPV, MH, and $\theta_{13}$ should be discovered for as small as possible true values of $\stheta$. For the sake of simplicity, we choose  maximal CP violation $\deltacp=\pi/2$ for the true $\deltacp$.\footnote{Other, more technical versions, are choosing the ``typical value of $\deltacp$'' (the median of the distribution in $\deltacp$), corresponding to a fraction of $\deltacp$ of 50\%, or a different certain fraction of $\deltacp$.  At least for CPV, our choice corresponds to the most optimistic case.} We show in \figu{LvsE-th13} the discovery reach in $\stheta$ for  maximal CP violation, MH, and $\theta_{13}$ as a function of baseline and $E_\mu$. The contours show the reach in (true) $\stheta$ for which the different quantities will be discovered  at the $3 \sigma$ CL. This figure is to be compared to Figs.~5 and~6 of \Ref~\cite{Huber:2006wb} for the respective $\deltacp$ and an older version of the detector simulation. Here the qualitative features are  clearly recovered: The CPV discovery requires a $2 \, 500$~km to $5 \, 000$~km baseline and $E_\mu$ above about 12~GeV. Note that degeneracies are typically unproblematic for this choice of $\deltacp$, whereas for $\deltacp=3 \pi/2$, a second baseline may be required.  In addition, note that relatively low $E_\mu$ are allowed because of the low detection threshold.  For the MH discovery, baselines longer than $4 \, 000 \, \mathrm{km}$ and $E_\mu$ larger than about 10-12~GeV are needed, since  the MSW resonance energy of about 8~GeV is to be covered. Here even longer baselines are preferred for different values of $\deltacp$. For the $\theta_{13}$ discovery, we find an extremely wide baseline and energy range, giving the least constraints. However, note again that this result depends on the choice of $\deltacp$. In summary, the result of this optimization, qualitatively, points towards one baseline between $2 \, 500$ and $5 \, 000 \, \mathrm{km}$ for the CPV measurement and one very long baseline for the MH measurement, such as the magic baseline at $7 \, 500 \, \mathrm{km}$ useful for degeneracy resolution (see \Ref~\cite{Huber:2006wb} for a more detailed discussion).
Because of the optimized detector, lower $E_\mu$ of down to 12~GeV may be possible. Below, we will discuss how this result changes for specific true values of $\stheta$ if all values of $\deltacp$ are considered.

\begin{figure}[tp]
 \centering
 \includegraphics[width=0.8\textwidth]{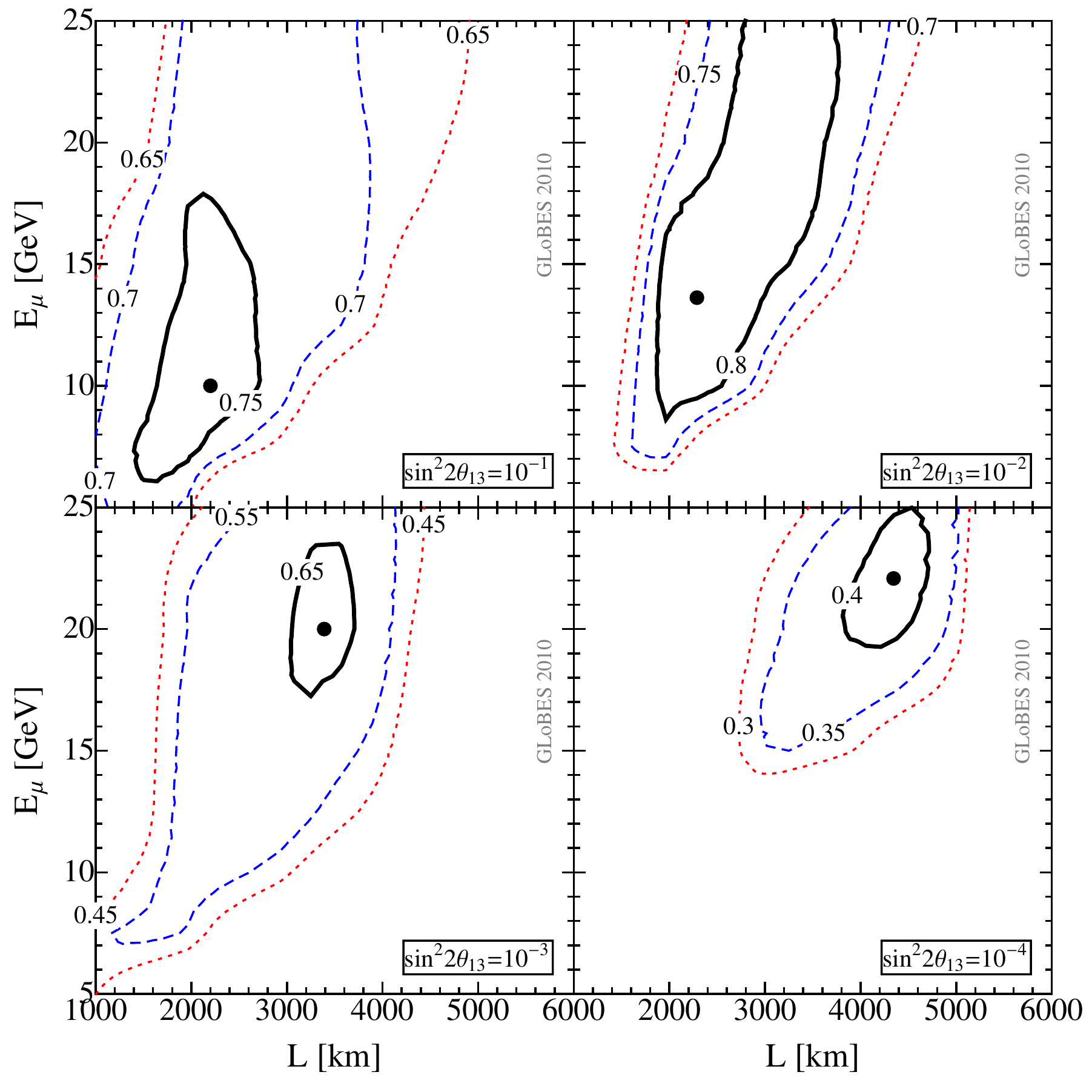}
 \mycaption{ \label{fig:LvsE}
 Fraction of $\deltacp$ for which CPV will be discovered ($3 \sigma$ CL) as a function of $L$ and $E_\mu$ for the single baseline Neutrino Factory.  The different panels correspond to different true values of $\stheta$, as given there. Here SF=1 is used with a 50 kt detector. The optimal performance is marked by a dot: (2200,10.00), (2288,13.62), (3390,20.00) and (4345,22.08) with regard to their best reaches of the fraction of $\deltacp$ at: 0.77, 0.84, 0.67 and 0.42.}
\end{figure}

From a different perspective, consider that the value of $\stheta$ is known, either from an earlier stage experiment or an earlier stage of the Neutrino Factory. In this case, as we have seen in the previous section, the MH discovery is typically not a problem (at least in combination with a longer baseline if $\stheta$ is small), and the most interesting question is the optimization of the fraction of $\deltacp$ for which CPV can be discovered. We first show in \figu{LvsE} the fraction of $\deltacp$ for which CPV will be discovered ($3 \sigma$ CL) as a function of $L$ and $E_\mu$ for the single baseline Neutrino Factory. The different panels correspond to different true values of $\stheta$, as given there. From this figure, it is obvious that the optimization strongly depends on the value of $\stheta$ chosen. For large $\stheta \simeq 10^{-1}$, shorter baselines and lower energies are preferred. Even $E_\mu$ as low as 5~GeV at the FNAL-Homestake baseline of about $1\, 300$~km is not far from optimal, which means that the MIND detector approaches the TASD performance of the low energy Neutrino Factory. Very interestingly, compared to earlier analyses without background migration, too high $E_\mu$ are in fact disfavored in the large $\stheta$ case.  Note that neither for the considered detector nor for the TASD, we find strong evidence supporting the ``bi-magic baseline'' in \Ref~\cite{Dighe:2010js}, see discussion in \App~\ref{app:bimagic}.
For the other extreme, $\stheta \simeq 10^{-4}$, baselines between $4 \, 000$ and $5 \, 000$~km are preferred with $E_\mu \simeq 20-25$~GeV, which corresponds more to the high energy Neutrino Factory, such as the IDS-NF baseline. Including the other two panels, the optimal region within each panel moves from the lower left on the plots to the upper right as the value of $\stheta$ decreases. This means that, depending on the choice of $\stheta$, the optimization results in the low energy Neutrino Factory, the high energy Neutrino Factory, or an intermediate scenario, and that the low and high energy Neutrino Factories are just two versions of the same experiment in different optimization regions. Of course, this discussion is somewhat hypothetical from the practical point of view, since either the next generation(s) of experiments will find $\stheta$ or not. If they find $\stheta$, the optimal parameters of the Neutrino Factory can be clearly predicted as a function of the detector response. The FNAL-Homestake low energy Neutrino Factory is one such possible setup for large enough $\stheta$ for the MIND detector. If they do not find $\stheta$, one may want to go for the IDS-NF high energy setup, which, in a way, represents the most aggressive but also inclusive option: This version of the Neutrino Factory is optimized for the worst case scenario.

\begin{figure}[tp]
 \centering
 \includegraphics[width=0.8\textwidth]{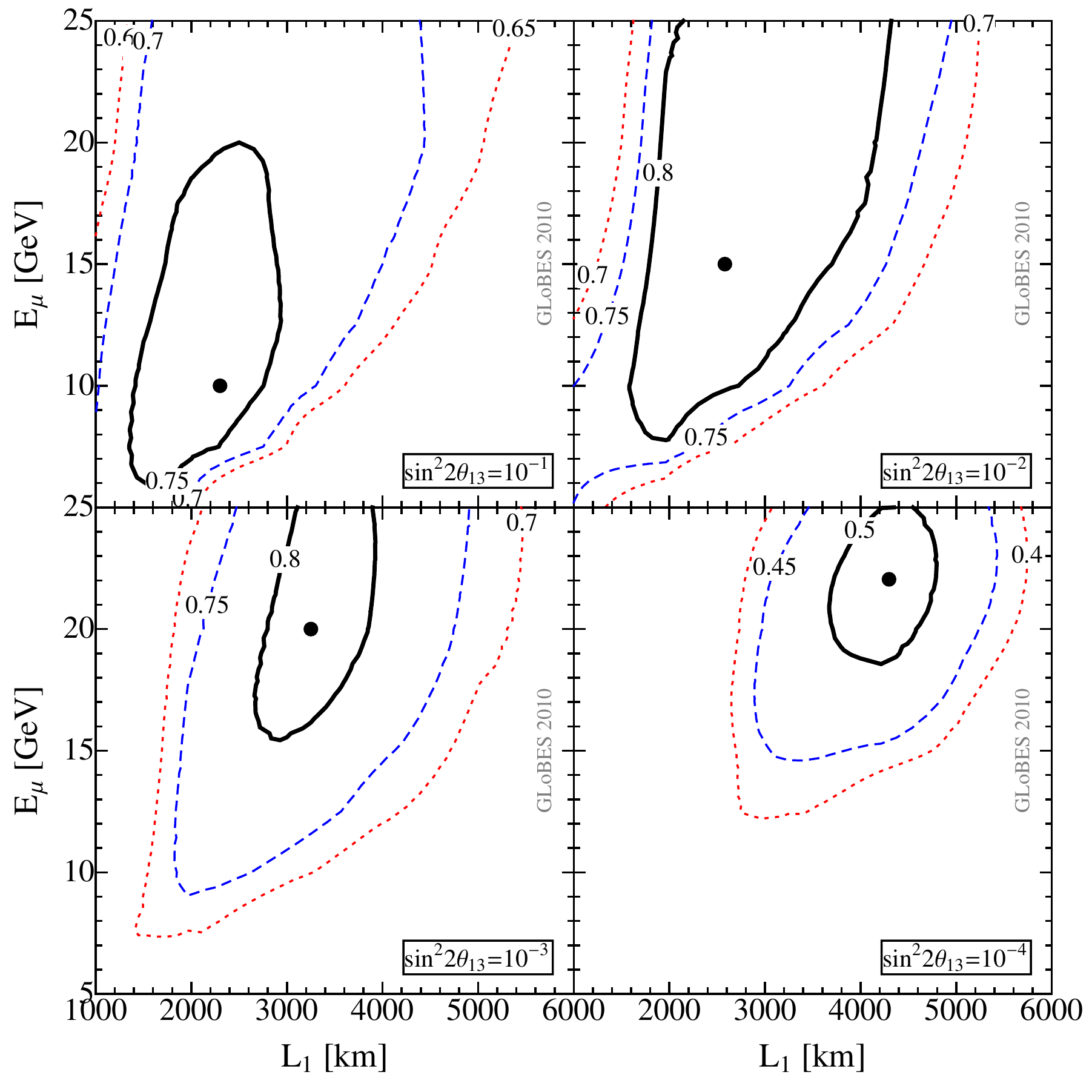}
 \mycaption{\label{fig:magic}Fraction of $\deltacp$ for which CPV will be discovered ($3 \sigma$ CL) as a function of $L_1$ and $E_\mu$ for the two-baseline Neutrino Factory, where $L_2 =7 \, 500 \, \mathrm{km}$ fixed.  The different panels correspond to different true values of $\stheta$, as given there. Here SF=1 is used and $E_\mu$ is assumed to be equal for both baselines. Two 50 kt detectors are used in the simulations. The optimal performance is marked by a dot: (2300,10.00), (2580,15.00), (3250,20.00) and (4297,22.05) followed by their best discovery reach of the fraction of $\deltacp$ at 0.77, 0.84, 0.81 and 0.51.}
\end{figure}

Apart from the single baseline, we show in \figu{magic} the combination with another fixed baseline $L_2 =7 \, 500 \, \mathrm{km}$ (in fact, there is typically very little dependence on the exact choice of the second baseline~\cite{Gandhi:2006gu}). Note that the muon energy is the same for both baselines. Comparing \figu{magic} with \figu{LvsE}, we find that the optimization of the short baseline hardly changes for very small and very large $\stheta$, whereas the possible baseline windows  for intermediate $\stheta$ (upper right and lower left panels) become somewhat broader. The energy optimization remains almost unaffected.  As far as the absolute performance is concerned, especially for $\stheta=10^{-3}$ and $\stheta=10^{-4}$, the fraction of $\deltacp$ increases because of the degeneracy resolution potential of the second baseline (which is not sensitive to $\deltacp$ itself by choosing exactly the magic baseline). For large values of $\stheta$, the second baseline is not required. This again reflects the correspondence to low and high energy Neutrino Factory: the low energy version is typically proposed with one baseline, the high energy version with two baselines.

\section{Earth geometry, and triangular shaped storage ring?}
\label{sec:site}

\begin{table}[t!]
 \begin{center}
 \begin{tabular}{||l|r|r|r|r||}
\hline\hline
  & \text{CERN} & \text{FNAL} & \text{J-PARC} &   \text{RAL} \\
 &(46.24,6.05) & (41.85,-88.28) & (36.47,140.57) & (51.57,-1.32) \\
\hline
 {\bf Asia:}  & & & & \\
 \text{CJPL} (28.15,101.71) & 7660 & 10420 & 3690 & 7840 \\
 \text{Kamioka} (36.14,137.24) & 8770 & 9160 & 300 & 8640 \\
 \text{YangYang} (37.77,128.89) & 8350 & 9300 & 1050 & 8270 \\
 \text{INO} (9.92,78.12) & 7360 & 11410 & 6570 & 7820 \\
\hline
  {\bf Europe:}  & & & & \\
 \text{LNGS} (42.37,13.44) & 730 & 7350 & 8840 & 1510 \\
 \text{Pyh{\"a}salmi} (63.68,25.98) & 2290 & 6630 & 7090 & 2080 \\
 \text{Slanic} (45.27,25.95) & 1540 & 7780 & 8150 & 2110 \\
 \text{Boulby} (54.56,-0.81) & 1050 & 5980 & 8480 & 340 \\
 \text{Canfranc} (42.76,-0.51) & 650 & 6550 & 9280 & 980 \\
 \text{Fr{\'e}jus} (45.20,6.67) & 130 & 6830 & 8900 & 920 \\
 \text{SUNLAB} (51.22,16.16) & 930 & 6980 & 8190 & 1210 \\
 \text{Umbria} (42.98,12.64) & 640 & 7280 & 8830 & 1420 \\
 \text{Gran Canaria} (28.39,-16.59) & 2780 & 6240 & 10570 & 2850 \\
\hline
 {\bf North America:}  & & & &  \\
 \text{Soudan} (47.82,-92.24) & 6590 & 730 & 8500 & 5900 \\
 \text{WIPP} (32.37,-104.23) & 8160 & 1760 & 8900 & 7540 \\
 \text{Homestake} (44.35,-103.77) & 7360 & 1290 & 8250 & 6690 \\
 \text{SNOLAB} (46.47,-81.19) & 6090 & 760 & 8950 & 5400 \\
 \text{Henderson} (39.77,-105.86) & 7750 & 1500 & 8410 & 7110 \\
 \text{Icicle Creek} (47.56,-120.78) & 7810 & 2610 & 7240 & 7160 \\
 \text{San Jacinto} (33.86,-116.56) & 8600 & 2610 & 8170 & 8000 \\
 \text{Kimballton} (37.37,-80.67) & 6580 & 820 & 9560 & 5950\\
\hline
\end{tabular}
\end{center}
\mycaption{\label{tab:labs} Here we show the baselines between the considered accelerator facilities (columns) and underground laboratories (rows) in kilometers. The latitude and longitude of each site is given in the brackets in degrees. The baselines are calculated using Mathematica with the International Terrestrial Reference Frame 2000, rounded to 10~km. All coordinates are consistently extracted from Google Maps~\cite{googlemap}.}
\end{table}

\begin{figure}[!thb]
 \centering
 \includegraphics[width=\textwidth]{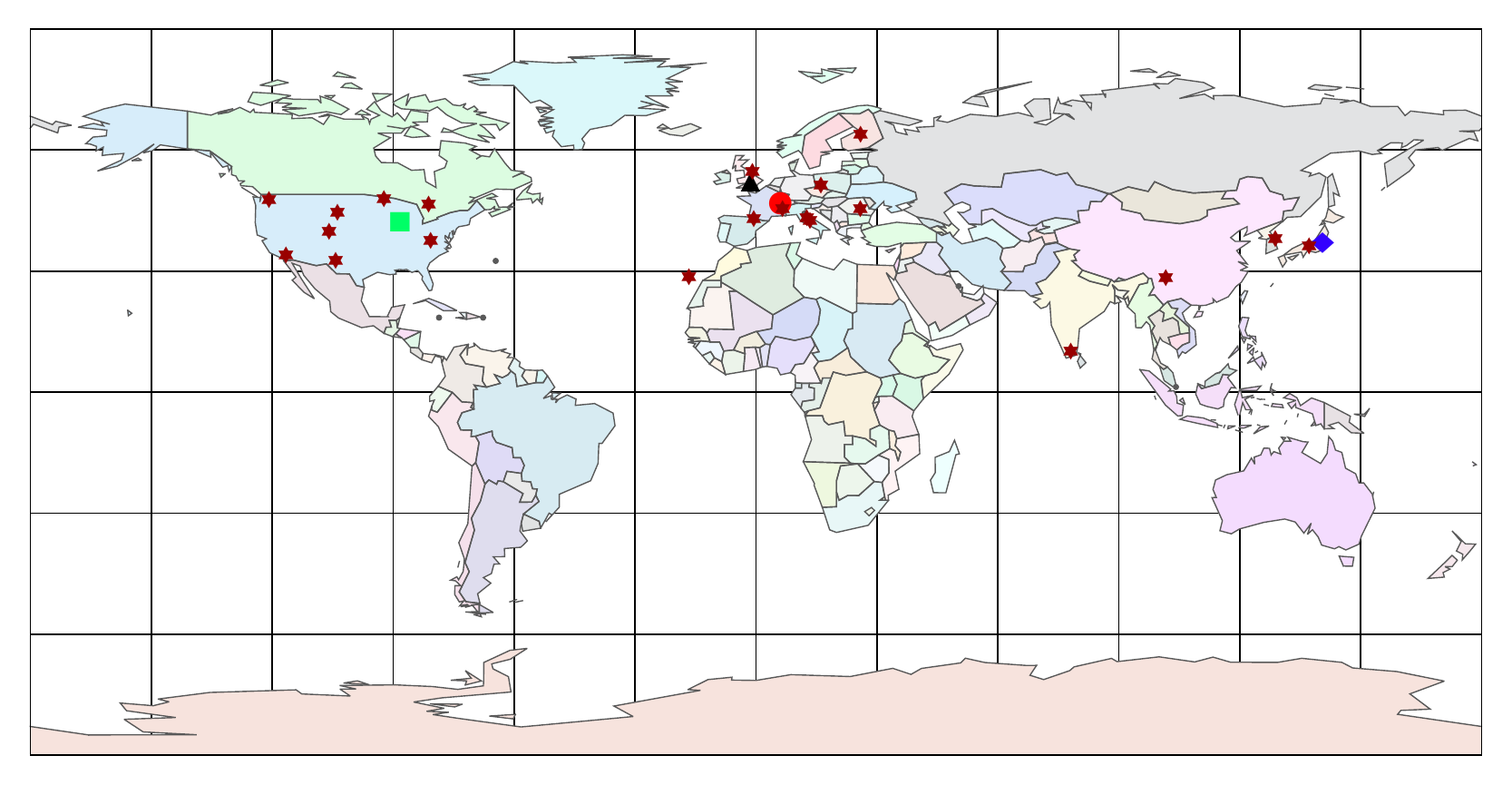}
 \mycaption{\label{fig:map}A world map including all potential sites of accelerators with large symbols and underground labs with stars.}
\end{figure}

In this section, we discuss the geometry aspects of specific sites for the high energy Neutrino Factory. The relevant questions for us are:
\begin{enumerate}
\item
 Can we find possible baseline combinations for the high energy Neutrino Factory for the large accelerator laboratories on different continents?
\item
 Would it be possible to use a single, triangular-shaped storage ring pointing towards both detector locations at the same time?
\end{enumerate}
We will quantify in the next section how specific baseline combinations translate into performance and optimization compared to the IDS-NF baseline parameters.

In order to address these purely geometric questions, we consider CERN, FNAL, J-PARC, and RAL as potential host laboratories for the Neutrino Factory.\footnote{Note that J-PARC is not very far away from KEK, for which the discussion would hardly change.} For the potential detector sites, we adopt the conservative point of view that significant rock overburden is needed. This assumptions and the anticipated timescale of the Neutrino Factory limits the choice of potential detector sites to currently investigated, or at least discussed, deep underground laboratories. We list the potential accelerator facilities and underground laboratories together with their locations and baselines between them in \Tab~\ref{tab:labs}; see \App~\ref{app:sites} for more details on the individual locations. The locations of laboratories and detector sites  on the Earth's surface can be found in \figu{map}.

 For CPV, the IDS-NF baseline has been 3000~km to 5000~km, based on the analysis in \Ref~\cite{Huber:2006wb}. This conclusion was obtained from the optimization of the $\theta_{13}$ reach, similar to \figu{LvsE-th13}, left panel. However, from
 \figu{LvsE} and \figu{magic} (see also Fig.~8 in \Ref~\cite{Tang:2009wp}), those shorter baselines are preferable if $\stheta$ turns out to be somewhat larger. Therefore, we allow for  $L_1\in(1500,5000)$~km for the high energy Neutrino Factory. For degeneracy resolution and the mass hierarchy measurement, $L_2$ should be close to the ``magic baseline''~\cite{Huber:2003ak} (\figu{LvsE-th13} is for one specific true value value of $\deltacp$), which can be see, for instance, in 
Fig.~5 of \Ref~\cite{Kopp:2008ds}. This location does not need to be exact. However, the baseline should not be to short, in order to allow matter effects to pile up and to suppress the CP violating terms, and not too long if too steep active storage rings legs should be avoided. We choose  $L_2\in(7000,8000)$ km as a reasonable range, see \Ref~\cite{Gandhi:2006gu}. 

\begin{figure}[!t]
 \centering
 \includegraphics[width=0.5\textwidth]{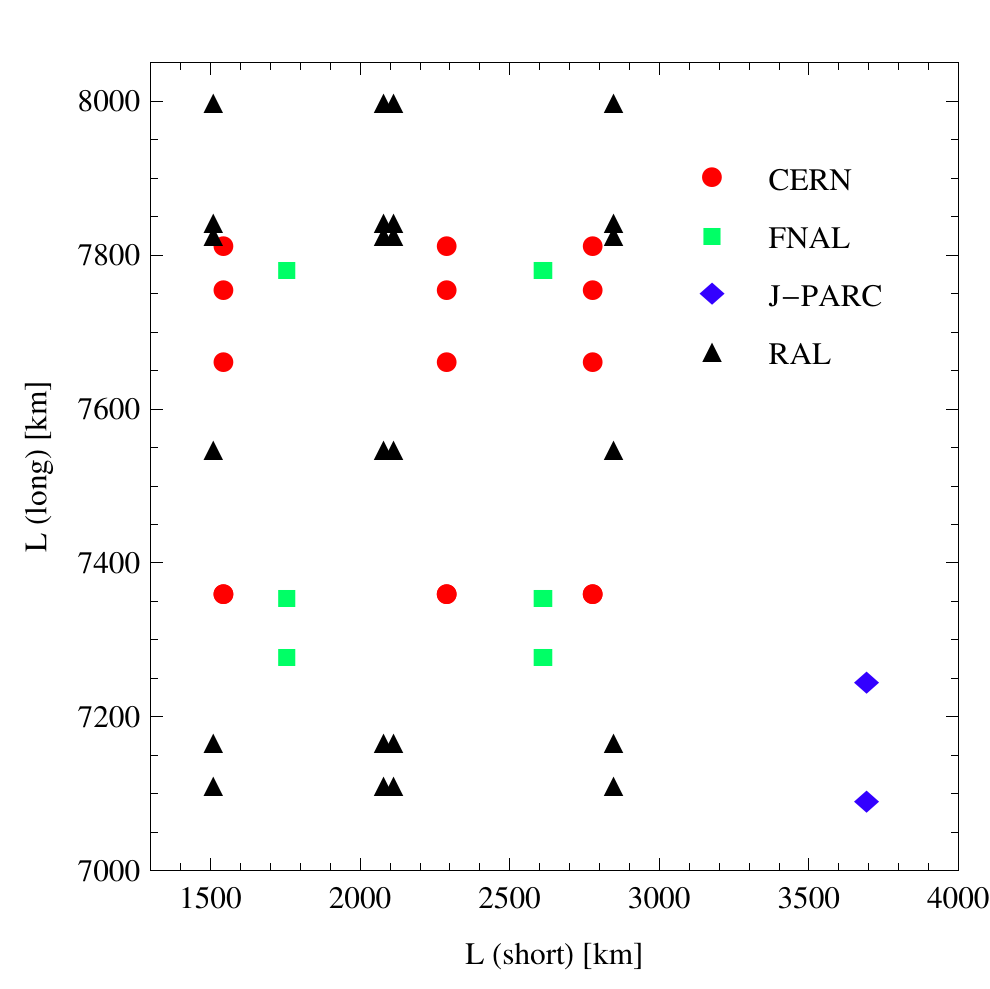}
 \mycaption{\label{fig:scatter} Baseline combinations from \Tab~\ref{tab:combine} shown in a two baseline plot. The different colors/symbols represent the different laboratories. }
\end{figure}

 As we can read off from \Tab~\ref{tab:labs}, there is a  very limited number of the short-baseline $L_1$ detector sites:
\begin{description}
 \item[CERN] {$L_1$: Pyh{\"a}salmi (Finland), Slanic-Prahova (Romania), Gran Canaria (Spain).}
 \item[FNAL] {$L_1$: WIPP, Henderson, Icicle Creek, San Jacinto.}
 \item[J-PARC] {$L_1$: CJPL.}
 \item[RAL] {$L_1$: LNGS (Italy), Pyh{\"a}salmi (Finland), Slanic-Prahova (Romania), Gran Canaria (Spain).}
\end{description}
We have not found any baseline between $4 \, 000$ and $5 \, 000$~km.
Obviously, we have plenty of options on the second baseline in comparison with the number of the first baseline. 
It may be noteworthy that CERN-INO and CERN-Homestake are exactly the same distances.

Now with these baseline windows for the short and long baselines, we can, for each laboratory, choose all possible combinations from \Tab~\ref{tab:labs}.
We show these in \figu{scatter}, with different shapes and colors for each laboratory, and we list them in \Tab~\ref{tab:combine}. Note that several qualitatively different baseline combinations in \figu{scatter} are marked (with the numbers from \Tab~\ref{tab:combine}), which we will discuss in the next section.
\begin{table}[p]
\begin{scriptsize}
 \begin{center}
  \begin{tabular}{||l|llclc|ccccc||}
\hline\hline
 \text{No.} & \text{Lab} & \text{Det1} & $L_1$ & \text{Det2} & $L_2$ & \text{Straight} & \text{Dead} &
   \text{Angle}[$^\circ$] & \text{SF} & \text{V-angle}[$^\circ$] \\
\hline
 \text{$\#$1} & \text{CERN} & \text{Pyh{\"a}salmi} & 2290 & \text{CJPL} & 7660 & 291 & 537 & 135 & 0.97 & 44 \\
 \text{$\#$2} & \text{CERN} & \text{Pyh{\"a}salmi} & 2290 & \text{INO} & 7360 & 306 & 507 & 112 & 1.02 & 35 \\
 \text{$\#$3} & \text{CERN} & \text{Pyh{\"a}salmi} & 2290 & \text{Homestake} & 7360 & 306 & 506 & 111 & 1.02 & 35 \\
 \text{$\#$4} & \text{CERN} & \text{Pyh{\"a}salmi} & 2290 & \text{Henderson} & 7750 & 308 & 503 & 110 & 1.03 & 38 \\
 \text{$\#$5} & \text{CERN} & \text{Pyh{\"a}salmi} & 2290 & \text{Icicle Creek} & 7810 & 299 & 520 & 120 & 1.00 & 39 \\
 \text{$\#$6} & \text{CERN} & \text{Slanic} & 1540 & \text{CJPL} & 7660 & 286 & 546 & 145 & 0.95 & 61 \\
 \text{$\#$7} & \text{CERN} & \text{Slanic} & 1540 & \text{INO} & 7360 & 284 & 550 & 151 & 0.95 & 75 \\
 \text{$\#$8} & \text{CERN} & \text{Slanic} & 1540 & \text{Homestake} & 7360 & 370 & 378 & 61 & 1.23 & 47 \\
 \text{$\#$9} & \text{CERN} & \text{Slanic} & 1540 & \text{Henderson} & 7750 & 370 & 378 & 61 & 1.23 & 50 \\
 \text{$\#$10} & \text{CERN} & \text{Slanic} & 1540 & \text{Icicle Creek} & 7810 & 354 & 411 & 71 & 1.18 & 45 \\
 \text{$\#$11} & \text{CERN} & \text{Gran Canaria} & 2780 & \text{CJPL} & 7660 & 391 & 336 & 51 & 1.30 & 77 \\
 \text{$\#$12} & \text{CERN} & \text{Gran Canaria} & 2780 & \text{INO} & 7360 & 368 & 383 & 63 & 1.23 & 52 \\
 \text{$\#$13} & \text{CERN} & \text{Gran Canaria} & 2780 & \text{Homestake} & 7360 & 313 & 493 & 104 & 1.04 & 36 \\
 \text{$\#$14} & \text{CERN} & \text{Gran Canaria} & 2780 & \text{Henderson} & 7750 & 311 & 497 & 106 & 1.04 & 38 \\
 \text{$\#$15} & \text{CERN} & \text{Gran Canaria} & 2780 & \text{Icicle Creek} & 7810 & 322 & 475 & 95 & 1.07 & 39 \\
\hline
 \text{$\#$16} & \text{FNAL} & \text{WIPP} & 1760 & \text{LNGS} & 7350 & 408 & 304 & 44 & 1.36 & 80 \\
 \text{$\#$17} & \text{FNAL} & \text{WIPP} & 1760 & \text{Slanic} & 7780 & 398 & 323 & 48 & 1.33 & 72 \\
 \text{$\#$18} & \text{FNAL} & \text{WIPP} & 1760 & \text{Umbria} & 7280 & 409 & 302 & 43 & 1.36 & 80 \\
 \text{$\#$19} & \text{FNAL} & \text{Icicle Creek} & 2610 & \text{LNGS} & 7350 & 345 & 430 & 77 & 1.15 & 42 \\
 \text{$\#$20} & \text{FNAL} & \text{Icicle Creek} & 2610 & \text{Slanic} & 7780 & 335 & 450 & 84 & 1.12 & 42 \\
 \text{$\#$21} & \text{FNAL} & \text{Icicle Creek} & 2610 & \text{Umbria} & 7280 & 345 & 429 & 77 & 1.15 & 42 \\
 \text{$\#$22} & \text{FNAL} & \text{San Jacinto} & 2610 & \text{LNGS} & 7350 & 384 & 352 & 55 & 1.28 & 61 \\
 \text{$\#$23} & \text{FNAL} & \text{San Jacinto} & 2610 & \text{Slanic} & 7780 & 371 & 376 & 61 & 1.24 & 57 \\
 \text{$\#$24} & \text{FNAL} & \text{San Jacinto} & 2610 & \text{Umbria} & 7280 & 384 & 350 & 54 & 1.28 & 61 \\
\hline
 \text{$\#$25} & \text{J-PARC} & \text{CJPL} & 3690 & \text{Pyh{\"a}salmi} & 7090 & 301 & 517 & 118 & 1.00 & 34 \\
 \text{$\#$26} & \text{J-PARC} & \text{CJPL} & 3690 & \text{Icicle Creek} & 7240 & 364 & 390 & 65 & 1.21 & 55 \\
\hline
 \text{$\#$27} & \text{RAL} & \text{LNGS} & 1510 & \text{CJPL} & 7840 & 304 & 511 & 114 & 1.01 & 39 \\
 \text{$\#$28} & \text{RAL} & \text{LNGS} & 1510 & \text{INO} & 7820 & 290 & 538 & 136 & 0.97 & 50 \\
 \text{$\#$29} & \text{RAL} & \text{LNGS} & 1510 & \text{WIPP} & 7540 & 408 & 302 & 43 & 1.36 & 82 \\
 \text{$\#$30} & \text{RAL} & \text{LNGS} & 1510 & \text{Henderson} & 7110 & 415 & 288 & 41 & 1.38 & 90 \\
 \text{$\#$31} & \text{RAL} & \text{LNGS} & 1510 & \text{Icicle Creek} & 7160 & 409 & 300 & 43 & 1.36 & 73 \\
 \text{$\#$32} & \text{RAL} & \text{LNGS} & 1510 & \text{San Jacinto} & 8000 & 403 & 313 & 46 & 1.34 & 87 \\
 \text{$\#$33} & \text{RAL} & \text{Pyh{\"a}salmi} & 2080 & \text{CJPL} & 7840 & 286 & 546 & 145 & 0.95 & 58 \\
 \text{$\#$34} & \text{RAL} & \text{Pyh{\"a}salmi} & 2080 & \text{INO} & 7820 & 296 & 526 & 125 & 0.99 & 41 \\
 \text{$\#$35} & \text{RAL} & \text{Pyh{\"a}salmi} & 2080 & \text{WIPP} & 7540 & 330 & 460 & 88 & 1.10 & 38 \\
 \text{$\#$36} & \text{RAL} & \text{Pyh{\"a}salmi} & 2080 & \text{Henderson} & 7110 & 324 & 470 & 93 & 1.08 & 35 \\
 \text{$\#$37} & \text{RAL} & \text{Pyh{\"a}salmi} & 2080 & \text{Icicle Creek} & 7160 & 312 & 494 & 105 & 1.04 & 34 \\
 \text{$\#$38} & \text{RAL} & \text{Pyh{\"a}salmi} & 2080 & \text{San Jacinto} & 8000 & 321 & 477 & 96 & 1.07 & 39 \\
 \text{$\#$39} & \text{RAL} & \text{Slanic} & 2110 & \text{CJPL} & 7840 & 290 & 539 & 137 & 0.97 & 48 \\
 \text{$\#$40} & \text{RAL} & \text{Slanic} & 2110 & \text{INO} & 7820 & 284 & 550 & 151 & 0.95 & 79 \\
 \text{$\#$41} & \text{RAL} & \text{Slanic} & 2110 & \text{WIPP} & 7540 & 393 & 333 & 50 & 1.31 & 67 \\
 \text{$\#$42} & \text{RAL} & \text{Slanic} & 2110 & \text{Henderson} & 7110 & 392 & 336 & 51 & 1.31 & 60 \\
 \text{$\#$43} & \text{RAL} & \text{Slanic} & 2110 & \text{Icicle Creek} & 7160 & 374 & 370 & 59 & 1.25 & 50 \\
 \text{$\#$44} & \text{RAL} & \text{Slanic} & 2110 & \text{San Jacinto} & 8000 & 380 & 359 & 56 & 1.27 & 62 \\
 \text{$\#$45} & \text{RAL} & \text{Gran Canaria} & 2850 & \text{CJPL} & 7840 & 377 & 366 & 58 & 1.26 & 63 \\
 \text{$\#$46} & \text{RAL} & \text{Gran Canaria} & 2850 & \text{INO} & 7820 & 347 & 425 & 75 & 1.16 & 46 \\
 \text{$\#$47} & \text{RAL} & \text{Gran Canaria} & 2850 & \text{WIPP} & 7540 & 318 & 483 & 99 & 1.06 & 37 \\
 \text{$\#$48} & \text{RAL} & \text{Gran Canaria} & 2850 & \text{Henderson} & 7110 & 323 & 472 & 94 & 1.08 & 36 \\
 \text{$\#$49} & \text{RAL} & \text{Gran Canaria} & 2850 & \text{Icicle Creek} & 7160 & 337 & 444 & 82 & 1.12 & 40 \\
 \text{$\#$50} & \text{RAL} & \text{Gran Canaria} & 2850 & \text{San Jacinto} & 8000 & 325 & 469 & 92 & 1.08 & 41 \\
\hline
 \text{$\#$51} & \text{FNAL} & \text{Homestake} & 1290 & \text{LNGS} & 7350 & 359 & 400 & 68 & 1.20 & 42 \\

\hline

\hline\hline
  \end{tabular}
 \end{center}
\end{scriptsize}
\vspace*{-0.5cm}
\mycaption{\label{tab:combine} Considered two-baseline combinations (see main text). The five right columns give the parameters of an isosceles triangle as storage ring: straight length (meters), dead length (meters), apex angle (degrees), scale factor SF (compared to two racetracks), angle to vertical (degrees).}
\end{table}
In addition to the setups with the above criteria, we have listed one option with FNAL-Homestake as first baseline (\#51). As we will demonstrate later, this baseline may be too short for the high energy Neutrino Factory.

Depending on the two baseline combination, it may be possible to use a triangular shaped storage ring instead of two racetracks. Here we follow the discussion in  \Ref~\cite{Berg:2008xx}, which the IDS-NF baseline setup with two storage rings is based on. The two racetrack-shaped storage rings are assumed to have a circumference of 1609~m. The active straights are about 600~m long, and, in each storage ring, $\mu^+$ and $\mu^-$ circulate in different directions. For a triangular shaped ring, probably two beam lines in the same tunnel are required to store $\mu^+$ and $\mu^-$ simultaneously. We assume that the circumference of the triangular ring, representative for the tunneling cost, is the same as for one racetrack, and we assume a (conservative) curvature radius $R_c$ of about 78~m for the curved sections. For the sake of simplicity, we only consider isosceles triangles with the same useful number of muon decays for the two far detectors. In the racetrack design of the IDS-NF baseline setup, $2.5 \cdot 10^{20}$ useful muons per year and polarity decay in each straight of each storage ring. For a triangular ring, all muons of one polarity can be injected into the same ring, leading to $5.0 \cdot 10^{20}$ useful muon decays per year and polarity over a straight length of 600~m, which corresponds to SF=2. Of course, due to the fixed circumference, the straights will be shorter than 600~m for the triangle, \ie, SF$<$2. Note that SF=1 corresponds to the same luminosity of the two designs. Obviously, if SF$>$1, the triangle is more efficient, possibly with a factor of two lower tunneling cost (because only one tunnel is needed). If SF$\lesssim$1, the loss of efficiency could be compensated by a slightly larger storage ring. For SF$\lesssim$0.5, the two racetracks are definitively the better option. 
Note that synchtrotron losses in the curved sections
of the storage rings and other beam losses are taken into
account for the racetrack-shaped geometry, yielding the
anticipated number of useful muon decays for the IDS-NF
baseline. We do not expect that these losses change drastically for a
triangular-shaped ring. For example, for the curved
sections, we have assumed the same curvature radius,
and the three curves sections add up to a circle just as for the two
curves sections of a racetrack-shaped ring. Since the circumference
of the triangular and racetrack shaped rings is the same, the
muons are exposed to the same curved section distance over
their lifetime.

\begin{figure}[!t]
 \centering
\begin{tabular}{ccc}
 \includegraphics[height=3cm]{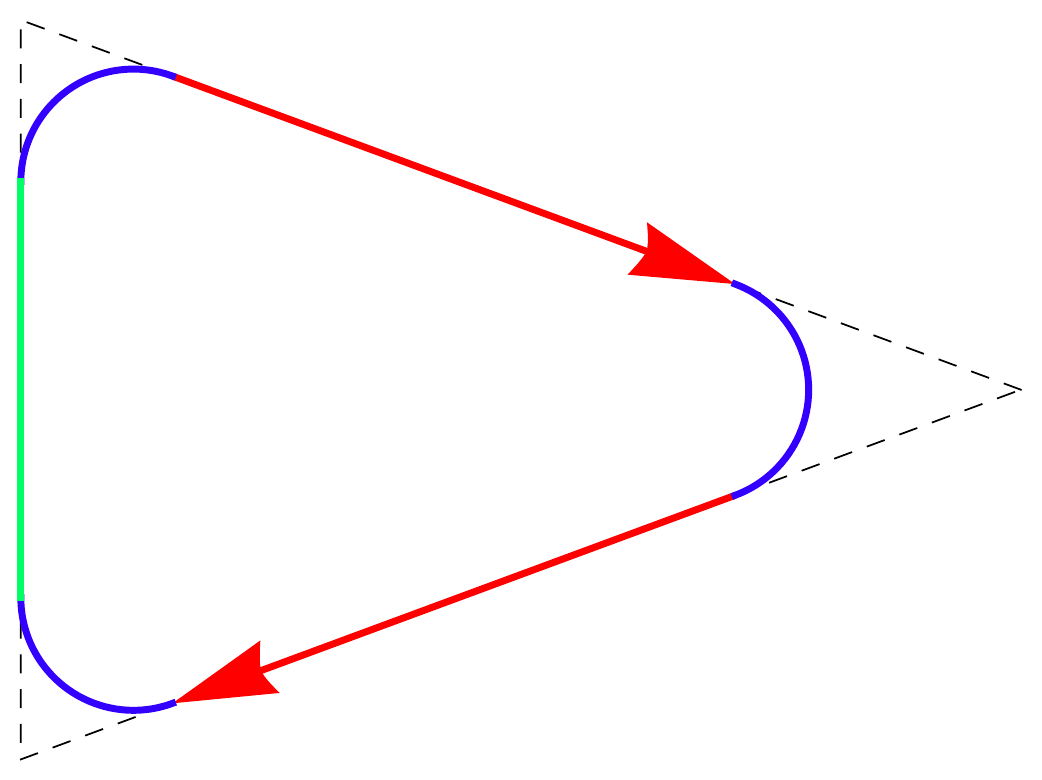} &
 \includegraphics[height=5cm]{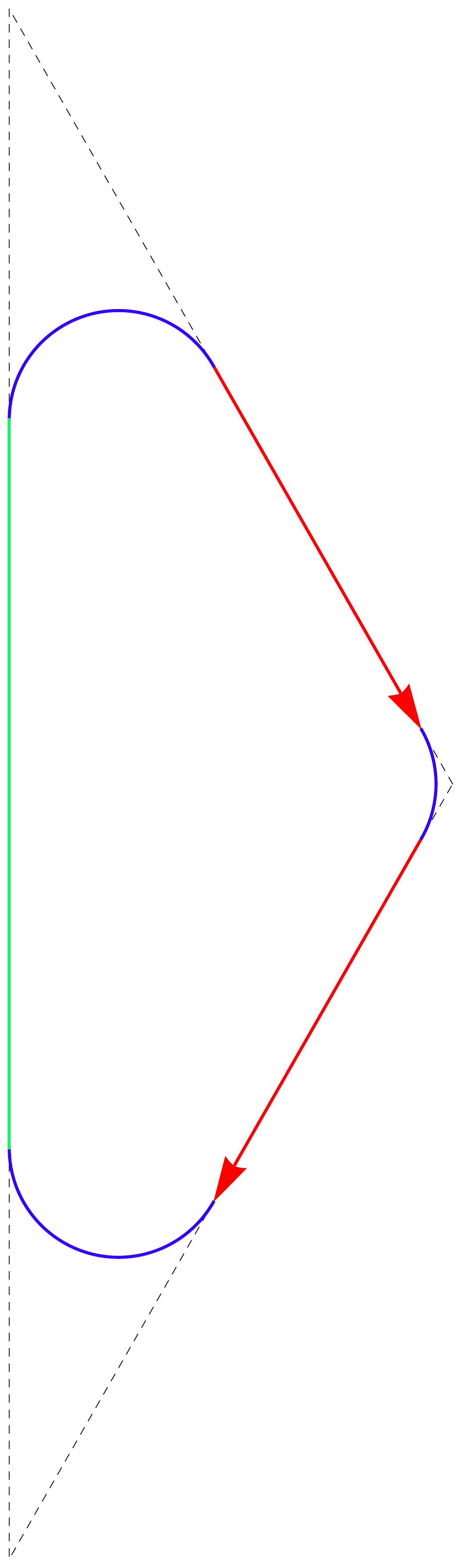} & \includegraphics[height=5.5cm]{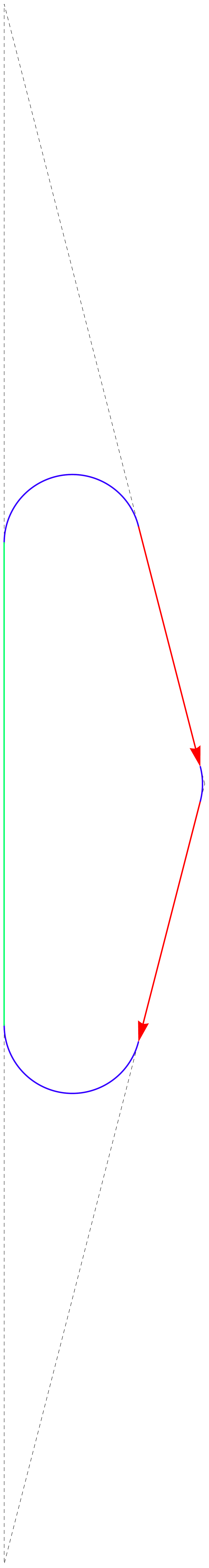} \\
\#30 & \#5 & \#40 \\
SF=1.38 & SF=1.00 & SF=0.95 
\end{tabular}
 \mycaption{\label{fig:triangles} Examples for three isosceles triangular-shaped rings (maximum efficiency, racetrack-like efficiency, and minimum efficiency, respectively). The numbers refer to \Tab~\ref{tab:combine}. }
\end{figure}

As the first observation, a triangular shaped ring can always be built if the circumference of the ring is larger than $2 \pi R_c$ ($R_c$: curvature radius curved section), which we have satisfied. This can be easily seen by the fact that in the smallest (extreme) case, the triangle with curved sections will collapse into a circle (with zero straight lengths). For larger triangles, the efficiency of the active legs (SF) may still be extremely small. However, we list in \Tab~\ref{tab:combine} the triangular geometry in terms of straight lengths, dead section length, apex angle, SF, and V-angle (angle between triangle plane and vertical), and it turns out that $0.95 \lesssim \mathrm{SF} \lesssim 1.38$. In the optimal case (\#30), the V-angle is $90^\circ$. This means that a triangle could be built for all of the considered options. We show three examples for maximum efficiency, racetrack-like efficiency, and minimum efficiency, respectively, in \figu{triangles}, where also the numbers from \Tab~\ref{tab:combine} are given. In the extreme cases, the triangle resembles a racetrack with either a very short or very long dead section. In the worst case, if the two detector locations are quite aligned, less then 50\% of the useful muon decays over the whole ring can be used. However, the factor of two higher muon injection rate compensates for that. 

In this discussion we have ignored how deep the tunnels would be and that two racetracks have other advantages. For instance, if one racetrack or one detector needs maintenance, all muons can be injected into the other storage ring without loss of performance integrated over the whole operation time. However, this discussion is interesting from a different perspective: Earlier in \Sec~\ref{subsec:performance}, we have shown that a single baseline operation may be more beneficial in parts of the parameter space, where one of the reasons is a factor of two gain in exposure compared to the operation of two racetracks. However, if a triangular ring is built, the argument changes. In \figu{detector}, the 100kt+50kt option has an exposure of 150~kt*1 (SF)=150~kt and the 100kt option an exposure of 100~kt*2 (SF)=200~kt . For the triangle, one in the most optimistic case for the 100kt+50kt option has an exposure of 150~kt*1.38 (SF)$\simeq$200~kt, which is the same as for the one baseline case -- at a much better sensitivity, and with the same storage ring circumference. From a different point of view, one has the performance depicted by 100kt+50kt in \figu{detector} in that case already with two 72~kt and 36~kt detectors.
Thus, for small $\stheta$ and proper detector sites, the triangle may finally be the better choice. 
Note that in the following, unless noted otherwise, we do not use the SF from \Tab~\ref{tab:combine}, but use SF=1 instead (two racetracks).

In summary, we have demonstrated that reasonable pairs of detector locations can be found for the considered accelerator laboratories. We have stated that one could always use a triangular-shaped storage ring with a similar efficiency as two racetracks from purely geometrical arguments, and that  the efficiency varies at about 40\% among the different options.

\section{Site-specific performance and energy optimization}
\label{sec:siteperf}

Here we discuss the performance of site-specific setups, as well as the optimization of $E_\mu$ for specific sites. Because of the large number of options considered, we only show examples in this section, whereas the plots for all discussed sites can be found in \App~\ref{app:baseline-combinations}.

\begin{figure}[tp]
 \centering
\includegraphics[width=1.0\textwidth]{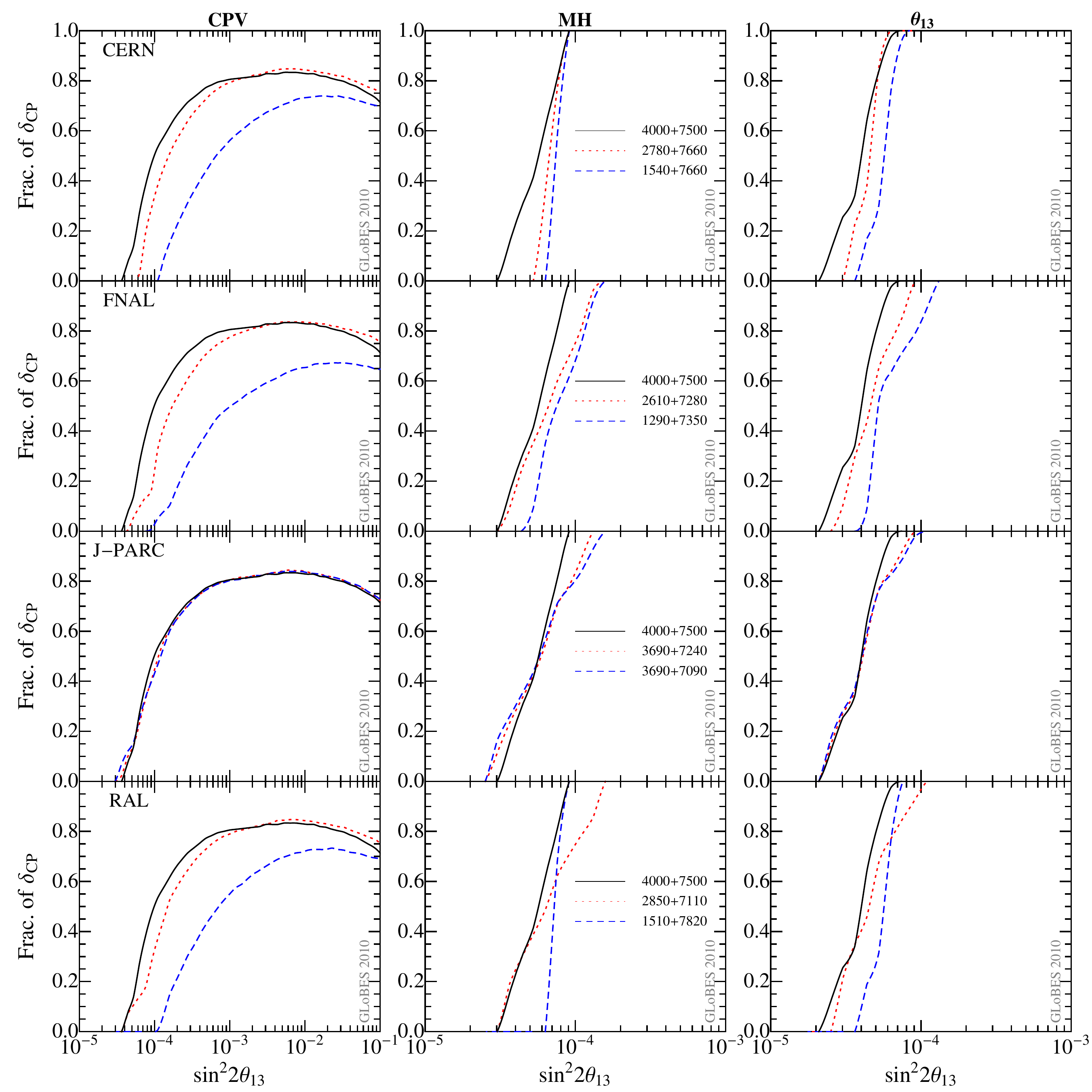}
 \mycaption{\label{fig:Sample} The discovery reach of CPV, MH, and $\theta_{13}$ ($3 \sigma$, in columns) for selected baseline combinations for different accelerator laboratories (in rows). In all panels, the curves for the IDS reference combination $4 \, 000$~km+$7 \, 500$~km with new-NF is shown for reference. Here $E_\mu$ is fixed to 25~GeV with two 50 kt detectors, and SF=1 is used in all cases (two racetrack-shaped storage rings).}
\end{figure}

Let us first of all quantify the performance in comparison to the IDS-NF baseline combination $4 \, 000$~km+$7 \, 500$~km at SF=1. Therefore, we show in \figu{Sample} the discovery reach for CPV, MH, and $\theta_{13}$ discovery ($3 \sigma$) for a number of qualitatively different selected baseline combinations for different accelerator laboratories (in rows). For each laboratory, we have chosen an example roughly representing the best case and an example close to the worst case for the chosen $E_\mu=25 \, \mathrm{GeV}$, as well as we show the IDS reference values. Here two racetrack-shaped storage rings (SF=1) are assumed. In all accelerator cases for CPV, options can be found which perform better than the IDS combination if $\stheta \gtrsim 10^{-2}$, because large values of $\stheta$ prefer shorter CPV baselines, as discussed earlier. In these cases, even a single baseline option with a lower $E_\mu$ could be preferable. For $10^{-3} \lesssim \stheta \lesssim 10^{-2}$, options close to the IDS performance can be easily identified.  For $\stheta \ll 10^{-3}$, the IDS combination can roughly be matched, but the sensitivity cannot be exceeded, at least not with the racetrack-shaped storage rings. The reason is that we do not use any baselines close to, or exceeding $4 \, 000$~km. Because of  the absence of potential detector sites, one may want to study either alternative locations, or the possibility to use MIND close to the surface. In this case, the long baseline may actually help for background suppression, since neutrinos from directions close to the beam have to travel through a significant amount of rock then. In the worst case scenarios, significant sensitivity losses may have to be taken into account, especially if not long enough CPV baselines are used. The MH discovery, on the other hand, is driven by the long baseline, but also benefits from a longer short baseline. For the optimal options, the IDS sensitivity can be matched, although it may be a bit different as a function of $\deltacp$.  Similar results are obtained for the $\theta_{13}$ discovery reach.

In \App~\ref{app:baseline-combinations}, we show the performances of the discussed combinations (\#1 to \#50) from \Tab~\ref{tab:combine}. In each case, we compare the performances with racetracks (SF=1, dotted curves), triangular-shaped geometry (SF from \Tab~\ref{tab:combine}, dashed curves), and IDS combination (SF=1, solid curves). Here we just mention some of the most interesting options from these figures, especially those with excellent sensitivity for $\stheta \lesssim 10^{-2}$ which can be further improved by a triangular shaped ring. Here CERN or RAL to Gran Canaria and to CJPL or INO are interesting options with a significant sensitivity gain and good absolute performance. In addition, J-PARC to CJPL and Icicle Creek is, in fact, the only option we find which can exceed the IDS reference performance for small $\stheta$. For options with shorter baselines, the performance also improves significantly in many cases by using a triangular geometry, but that cannot compensate for the baseline choice. In no case, the performance is significantly worse using a triangle.

\begin{figure}[tp]
 \centering
 \includegraphics[width=0.8\textwidth]{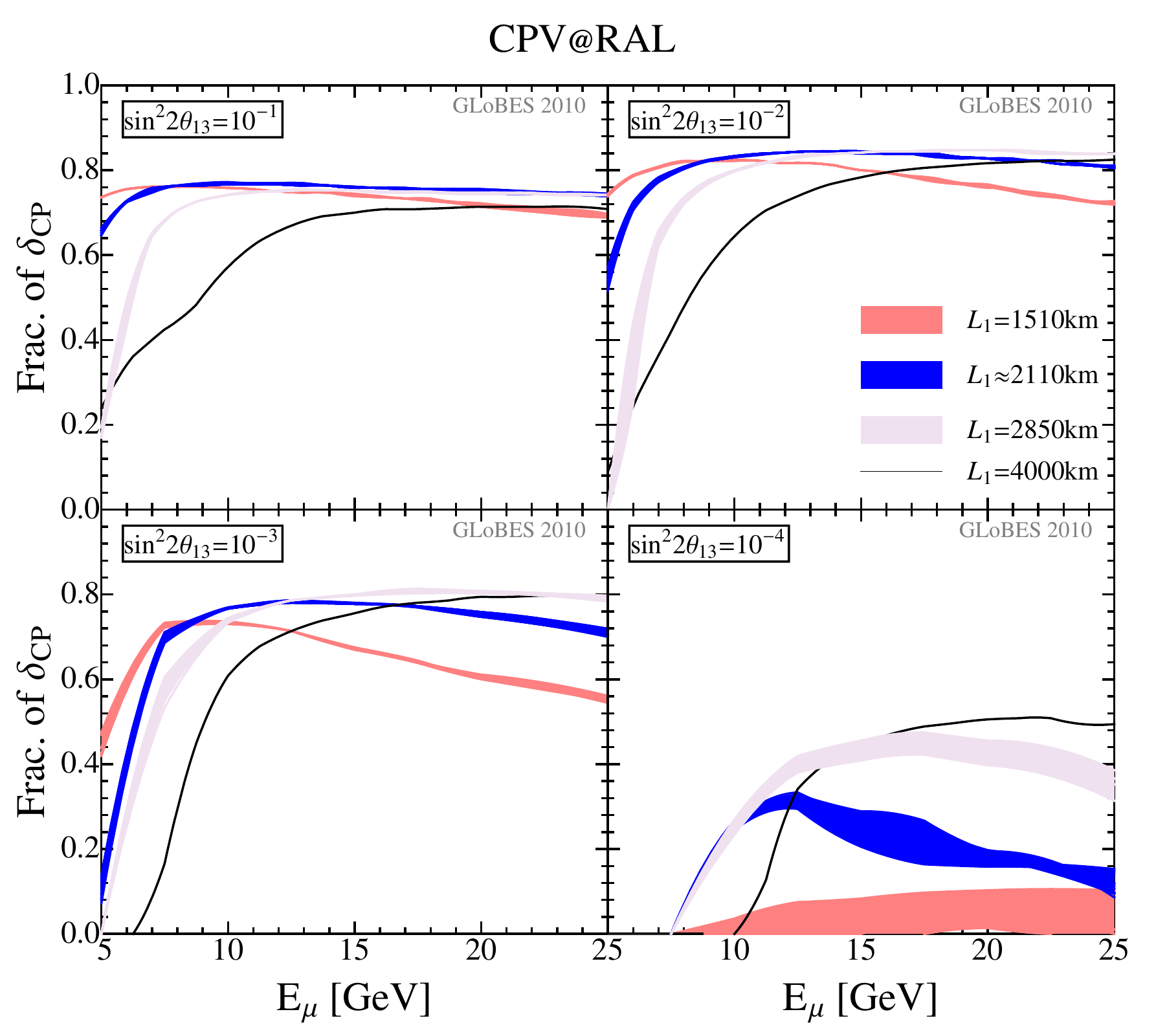}
 \mycaption{CPV discovery reach at RAL as a function of $E_\mu$ for all RAL-options in \Tab~\ref{tab:combine} at the $3 \sigma$ CL for different values of (true) $\stheta$ (as given in the panels).  Here we assume SF=1.0 and two 50 kt detectors. We group the different baseline combinations according to the shorter baseline $L_1$, as shown in the legend. The bands basically reflect the variation of the second baseline. The IDS-NF baseline combination is shown by the black curves with $L_1=4000$~km and $L_2=7500$~km.}
 \label{fig:RAL}
\end{figure}

Another interesting question is the optimization of $E_\mu$ for specific two baseline setups. Remember that the two baseline Neutrino Factory is mostly relevant for small $\stheta$. Here we choose the combinations from RAL as example, see \figu{RAL}, the other laboratories are shown in \App~\ref{app:baseline-combinations}. Basically, we can identify three different sets of curves in that figure, which correspond to the three different CPV baselines in \Tab~\ref{tab:combine}:  (a) $L_1=1 \, 510$~km, (b) $L_1 \simeq 2 \, 100$~km (two different ones), (c) $L_1 =2 \, 848$~km. Depending on the value of $\stheta$ and $E_\mu$, one of these three sets performs best: below about 7-8~GeV (depending on $\stheta$), (a) is best, between about 8 and 12~GeV (b) is best, and above 10-14 GeV, (c) is best. This results more or less reproduces the green-field optimization. Note that for $\stheta \lesssim 10^{-2}$, the case for which the two baseline Neutrino Factory is the relevant choice, the long CPV baseline options are better in terms of absolute performance, provided that $E_\mu$ is high enough. In addition, note that the IDS reference prefers $E_\mu \gtrsim 20 \, \mathrm{GeV}$ in all cases, where the discovery reaches saturate. To summarize, the optimal muon energy does not only depend on $\stheta$, but also on the specific two baseline combination. However, in many cases, the performance saturates at about 12-15~GeV (see, \eg,   $\stheta=10^{-4}$), and in some cases may even decrease for too high $E_\mu$. The IDS-NF baseline choice $E_\mu = 25 \, \mathrm{GeV}$ can be understood as an aggressive option from the current point of view. However, note that for any given baseline combination, the optimization of $E_\mu$ can be easily performed. From the machine point of view, it should be easy to ``down-grade'' the setup then.

\section{Summary and conclusions}
\label{sec:conclusion}

\begin{figure}[tp]
 \centering
 \includegraphics[width=1.0\textwidth]{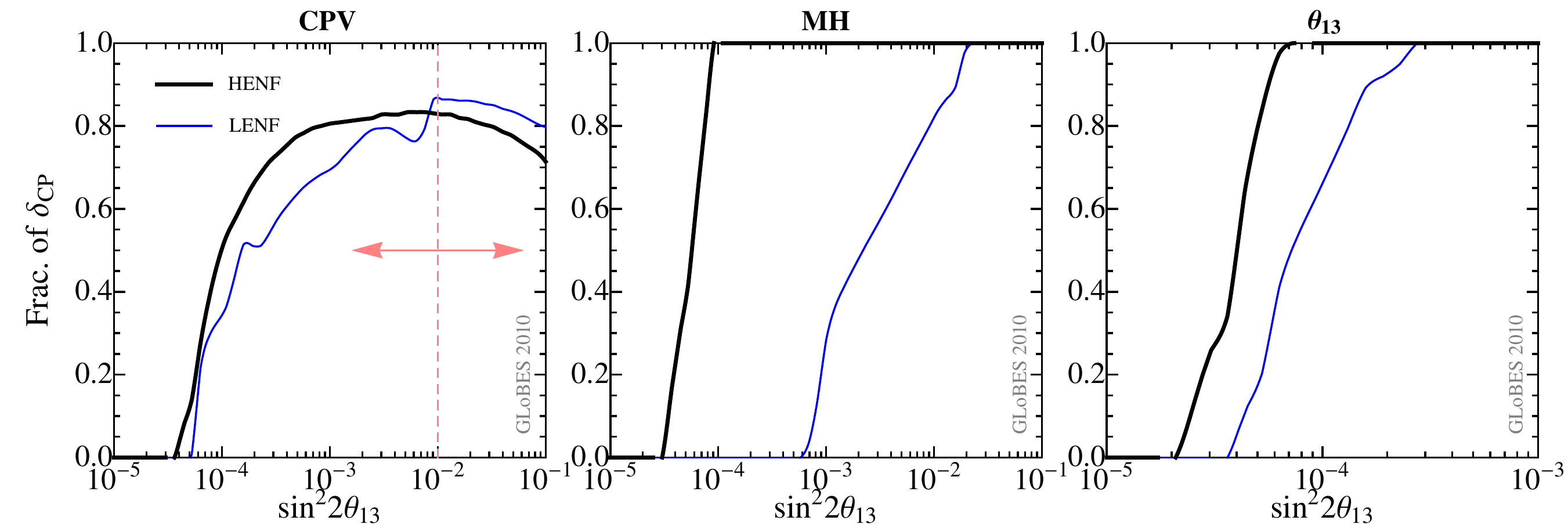}
 \mycaption{A comparison of the discovery reaches of CPV, MH, and $\theta_{13}$ at the $3\sigma$ CL between a low energy single baseline neutrino factory (LENF) with $E_\mu=10$ GeV and $L=2000 \, \mathrm{km}$ (\cf, \figu{LvsE}, upper left panel), and a high energy two baseline neutrino factory (HENF) with $E_\mu=25$~GeV, $L_1=4000 \, \mathrm{km}$, and $L_2=7500 \, \mathrm{km}$ (\cf, \figu{magic}, lower right panel).}
 \label{fig:LENF}
\end{figure}

In this study, we have revisited the optimization of the Neutrino
Factory based on the most up-to-date analysis of the MIND detector
using migration matrices. We have also considered possible backgrounds
from taus, which come from $\nu_\tau$ charged current interactions
in the detector, and which practically cannot be distinguished from
muons. We have found that the resulting backgrounds do not have a
visible impact on the CPV, MH, and $\theta_{13}$ discovery reaches.  A
more refined discussion will require, however, a consistent treatment
of all migration matrices.

Although the optimization of the Neutrino Factory does generically not
change with the new detector simulation, there are a number of
interesting observations. The lower threshold and higher efficiencies
compared to earlier simulations imply that the MIND detector
characteristics are getting more similar to the characteristics of the
detectors proposed for the low energy Neutrino Factory (\eg, a
magnetized TASD).  We could demonstrate that we recover the
$L$-$E_\mu$-optimization of the low energy Neutrino Factory for large
$\stheta$: In this case, a single baseline Neutrino Factory with
$E_\mu$ as low as 5~GeV and a baseline as short as FNAL-Homestake
(about $1 \, 300$~km) might be sufficient.  For small $\stheta <
10^{-2}$, however, we find that a two baseline Neutrino Factory with
one baseline between about $2 \, 500$~km and $5 \, 000$~km, the other
one at about the magic baseline $7 \, 500$~km is ideal. This recovers
the results from \Ref~\cite{Huber:2006wb}.  We summarize this in
\figu{LENF}, where we compare the performance of the optimal single
baseline low energy Neutrino Factory with the optimal two baseline
high energy Neutrino Factory for the same MIND detector.  One can
clearly see that for $\stheta \gtrsim 10^{-2}$ the low energy version
can perform all of the required measurements, whereas for smaller
values the high energy Neutrino Factory is clearly better.  This means
that this different optimization would be sufficient to compensate for
the relative deterioration of performance at large $\stheta$ observed
in the traditional high-energy Neutrino Factory.  If $\stheta$ was
known, the shorter (CPV) baseline could even be optimized: The larger
$\stheta$ was, the shorter CPV baselines would be preferred in the
mentioned baseline window.  The next generation of experiments will
tell us if $\stheta \gtrsim 10^{-2}$ or smaller, see, \eg,
\Ref~\cite{Huber:2009cw}, therefore, we can optimize for the large
$\stheta>0.01$ case. Note that a more refined optimization depending
on the size of $\stheta$ may be possible for a staged Neutrino Factory
approach, as it is illustrated in \Ref~\cite{Tang:2009wp}.

Apart from the optimization of the green-field Neutrino Factory, we
have performed a site-specific analysis for the high energy Neutrino
Factory, assuming that it requires two baselines. We have considered
four different accelerator laboratories on three different continents
(CERN, FNAL, J-PARC, RAL) and a number of potential detector locations
in suggested underground laboratories. We have found that in all cases
plausible baseline combinations can be found. However, for small
$\stheta$, where a baseline between $2 \,500$ and $5\, 000$~km is
preferred for CPV, we only found one possible baseline: J-PARC to CJPL
(China). Therefore, we propose that possible underground sites for
this baseline window should be investigated. In addition, we propose
to study the MIND performance on the surface, since surface operation
would greatly facilitate site and baseline selection.

We have also investigated the possibility to use a triangular-shaped
muon storage ring compared to two racetracks, where the efficiency is
a function of the Earth geometry and the chosen source and detector
locations. We have first of all shown that solely based on geometry a
triangular ring could be used in either case, without significant loss
of luminosity. Then we have identified a number of baseline
combinations with reasonable baseline lengths for which the triangle
would be especially interesting: CERN or RAL to Gran Canaria and to CJPL 
or INO.  In addition, J-PARC to CJPL and Icicle Creek is, in fact, the only 
option we found which can exceed the IDS reference performance for small $\stheta$.
 We have also pointed out that using a triangular-shaped
ring, the decision between one and two baselines does not emerge, and
that, from the physics point of view, the two baseline combination is
more efficient for the same exposure.

As far as the optimization $E_\mu$ is concerned, the feature of the
new detector simulation that the backgrounds are typically
reconstructed at lower energies and that the threshold is lower leads
to new insights. Especially for the high energy Neutrino Factory in
the context of specific baseline combinations, the CPV performance
in some cases saturates at lower muon energies than $25 \, \mathrm{GeV}$. 
Although the current IDS-NF
baseline setup with $E_\mu=25 \, \mathrm{GeV}$ is still optimal for
the optimal baseline combination, and high $E_\mu$ typically do not
harm for small $\stheta$, smaller energies may be preferred for
specific sites.

We conclude that the low energy and high energy Neutrino Factory
should not be regarded as separate options. Let us emphasize that the
optimization, for instance, of $E_\mu$ is a function of
$\stheta$, the detector response, and the specific sites chosen for
detector and accelerators. Therefore, the IDS-NF baseline with
$E_\mu=25$~GeV should be understood as most conservative choice, which
can be downgraded in specific scenarios/for specific detectors. From
the machine point of view, we recommend to choose splitting points
between the different accelerator components at about 5 and 12~GeV,
which will allow for $E_\mu=5$, 12, or 25~GeV. The final choice has to
be made based on the knowledge on $\stheta$ at the time of decision,
the choice of the detector, and the specific site.

\subsubsection*{Acknowledgments}

We thank Alain Blondel, Anselmo Cervera, Pilar Coloma, Andrea
Donini, and Paul
Soler for insightful discussions and thank Andrew Liang and Davide Meloni
for providing the latest data files of the migration matrices. JT is
also indebted to Xiaobo Huang for help with the usage of ROOT.

This work has been supported by the Emmy Noether program of Deutsche
Forschungsgemeinschaft (DFG), contract no. WI 2639/2-1 [J.T, W.W.], by the
DFG-funded research training group  1147 ``Theoretical astrophysics and
particle physics'' [J.T.], and by the European Union under the
European Commission Framework Programme~07 Design Study EUROnu,
Project 212372. This work has also been supported by the U.S. Department of
Energy under award number \protect{DE-SC0003915}. S.K.A acknowledges
in addition the support from the project Consolider-Ingenio CUP.

\begin{appendix}

\section{Locations of accelerator facilities and underground laboratories}
\label{app:sites}

We use Google Maps~\cite{googlemap} to find the exact locations 
(latitudes and longitudes) of considered accelerator facilities and underground laboratories.
In the following, we provide the details of the locations based on which the latitudes and longitudes have been obtained.

\subsection{Accelerator facilities}

\begin{itemize}

\item
CERN

Latitude  : 46.24$^\circ$ N \&
Longitude :  6.05$^\circ$ E

Route de Meyrin 385, 1217 Geneve, Schweiz, Switzerland.

\item
FNAL

Latitude  : 41.85$^\circ$ N \&
Longitude : 88.28$^\circ$ W

Center for Particle Astrophysics, Fermi National Accelerator Laboratory, P.O. Box 500, Kirk Road and Pine Street, Batavia, Illinois 60510-0500 USA.

\item
J-PARC

Latitude  :  36.47$^\circ$ N \&
Longitude : 140.57$^\circ$ E

Tokai Village, Naka District, Ibaraki Prefecture, Japan.

\item
RAL

Latitude  :  51.57$^\circ$ N \&
Longitude :   1.32$^\circ$ W

Rutherford Appleton Laboratory, Harwell Science \& Innovation Campus, Didcot OX110QX, UK.

\end{itemize}   


\subsection{Underground facilities in USA}

\begin{itemize}

\item
Soudan

Latitude  :  47.82$^\circ$ N \&
Longitude :  92.24$^\circ$ W

Soudan Underground Lab, 30 1st Avenue, Soudan, MN 55782.

\item
WIPP

Latitude  :  32.37$^\circ$ N \&
Longitude : 104.23$^\circ$ W

The WIPP Experience Exhibit, U.S. Department of Energy, 4021 National Parks Highway, Carlsbad, 
New Mexico.

\item
Homestake

Latitude  :  44.35$^\circ$ N \&
Longitude : 103.77$^\circ$ W

Homestake Visitor Center, 160 West Main Street, Lead, SD 57754-1362.

\item
SNOLAB

Latitude  :  46.47$^\circ$ N \&
Longitude :  81.19$^\circ$ W

SNOLAB, Greater Sudbury, Ontario, Canada.

\item
Henderson

Latitude  :  39.77$^\circ$ N \&
Longitude : 105.86$^\circ$ W

Henderson Mine and Mill, 1746 County Road 202 Empire, CO 80438.

\item
Icicle Creek

Latitude  :  47.56$^\circ$ N \&
Longitude : 120.78$^\circ$ W

Bridge Creek Campground, Leavenworth, Washington 98826.

\item
San Jacinto

Latitude  :  33.86$^\circ$ N \&
Longitude : 116.56$^\circ$ W

Mt San Jacinto State Park, 1 Tramway Road, Palm Springs, CA 92262-1827.

\item
Kimballton

Latitude  :  37.37$^\circ$ N \&
Longitude :  80.67$^\circ$ W

Kimballton, VA 24136.

\end{itemize}


\subsection{Underground facilities in Europe}

\begin{itemize}

\item
LNGS

Latitude  :  42.37$^\circ$ N \&
Longitude :  13.44$^\circ$ E

Istituto Nazionale Di Fisica Nucleare - Laboratori Nazionali Del Gran Sasso-Ufficio Amministrativo, Strada Statale 17 Bis, L'Aquila, AQ 67100, Italy.

\item
Pyh{\"a}salmi

Latitude  :  63.68$^\circ$ N \&
Longitude :  25.98$^\circ$ E

86800 Pyhajarvi municipality in the south of Oulu province, Finland.

\item
Slanic

Latitude  :  45.27$^\circ$ N \&
Longitude :  25.95$^\circ$ E

Largest Salt mine in Europe, Prahova, Slanic, Romania.

\item
Boulby

Latitude  :  54.56$^\circ$ N \&
Longitude :   0.81$^\circ$ W

Boulby Potash Mine located just southeast of the village of Boulby, on the northeast coast of the North Yorkshire Moors in Redcar and Cleveland, England.

\item
Canfranc

Latitude  :  42.76$^\circ$ N \&
Longitude :   0.51$^\circ$ W

Laboratorio Subterr{\'a}neo de Canfranc lies physically between New Road tunnel and Old Railway tunnel of Canfranc, Spain.

\item
Fr{\'e}jus

Latitude  :  45.20$^\circ$ N \&
Longitude :   6.67$^\circ$ E

Laboratoire souterrain de Modane, Carre Sciences, 73500 Modane, France.

\item
SUNLAB

Latitude  :  51.22$^\circ$ N \&
Longitude :  16.16$^\circ$ E

Polkowice-Sieroszowice mine near Wroclaw in Poland.

\item
Umbria

Latitude  :  42.98$^\circ$ N \&
Longitude :  12.64$^\circ$ E

Umbria, IT in Italy.

\item
Gran Canaria

Latitude  :  28.39$^\circ$ N \&
Longitude :  16.59$^\circ$ W

Gran Canaria in Spain.

\end{itemize}


\subsection{Underground facilities in Asia}

\begin{itemize}

\item
CJPL

Latitude  :   28.15$^\circ$ N \&
Longitude :  101.71$^\circ$ E

China JinPing Deep Underground Laboratory at Sichuan Province in China close to Jinping
mountain.

\item
Kamioka

Latitude  :   36.14$^\circ$ N \&
Longitude :  137.24$^\circ$ E

Kamioka is located underground in the Mozumi Mine of the Kamioka Mining and Smelting Co. near 
the Kamioka section of the city of Hida in Gifu Prefecture, Japan.

\item
YangYang

Latitude  :   37.77$^\circ$ N \&
Longitude :  128.89$^\circ$ E

YangYang underground laboratory (Y2L) is located at a depth of 700 m under an earth overburd
en in South Korea.

\item
INO

Latitude  :   9.92$^\circ$ N \&
Longitude :  78.12$^\circ$ E

Bodi West Hills Reserved Forest in Theni district of Tamil Nadu, India.

\end{itemize}

  
\section{Details for all considered two-baseline combinations}
\label{app:baseline-combinations}

\begin{figure}[tp]
 \centering
 \includegraphics[width=0.8\textwidth]{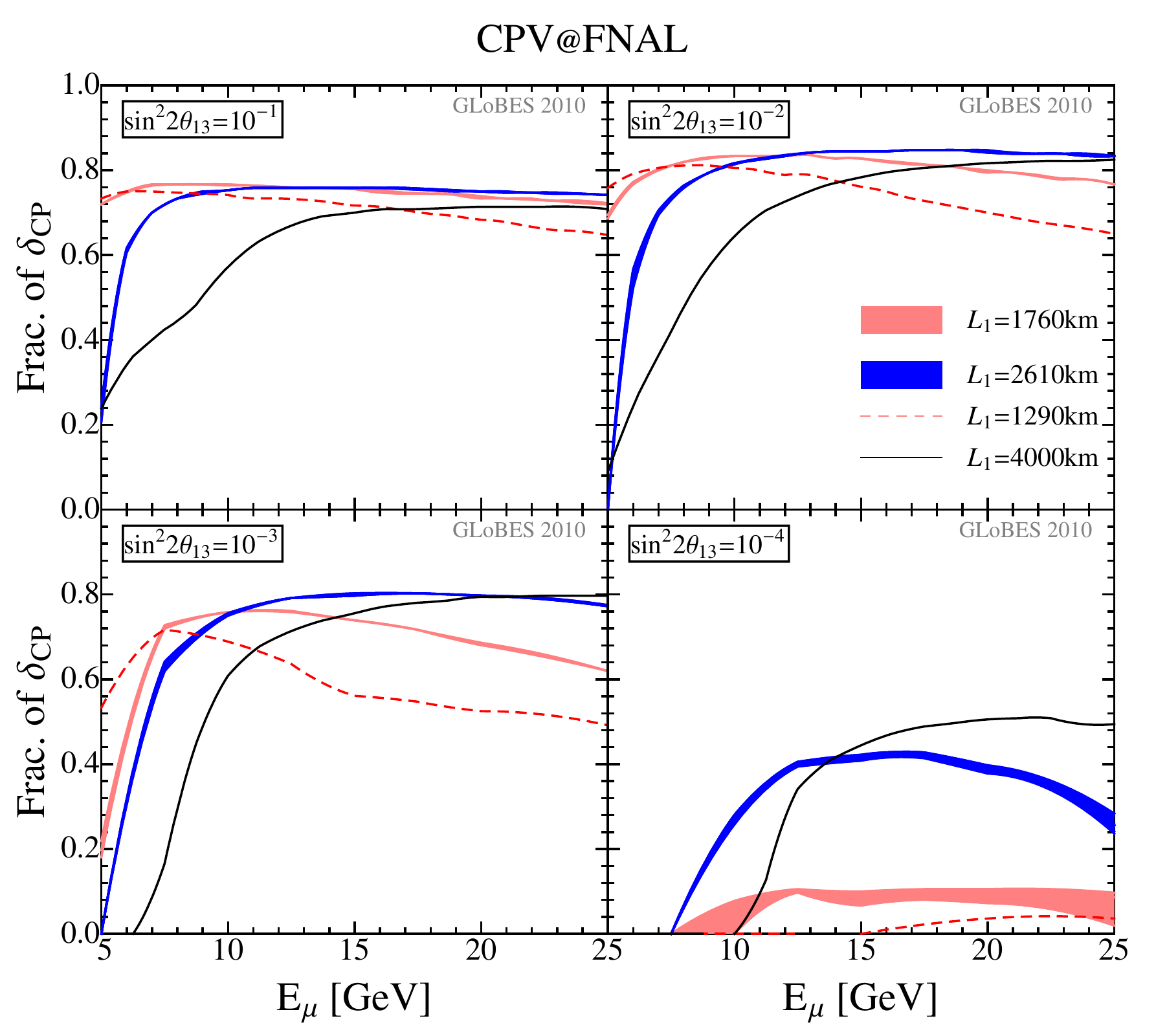}
 \mycaption{CPV discovery reach at FNAL as a function of $E_\mu$ for all FNAL-options in \Tab~\ref{tab:combine} at the $3 \sigma$ CL for different values of (true) $\stheta$ (as given in the panels).  Here we assume SF=1.0 and two 50 kt detectors. We group the different baseline combinations according to the shorter baseline $L_1$, as shown in the legend. The bands basically reflect the variation of the second baseline. The IDS-NF baseline combination is shown by the black curves with $L_1=4000$ km and $L_2=7500$~km.}
 \label{fig:FNAL}
\end{figure}

\begin{figure}[tp]
 \centering
 \includegraphics[width=0.8\textwidth]{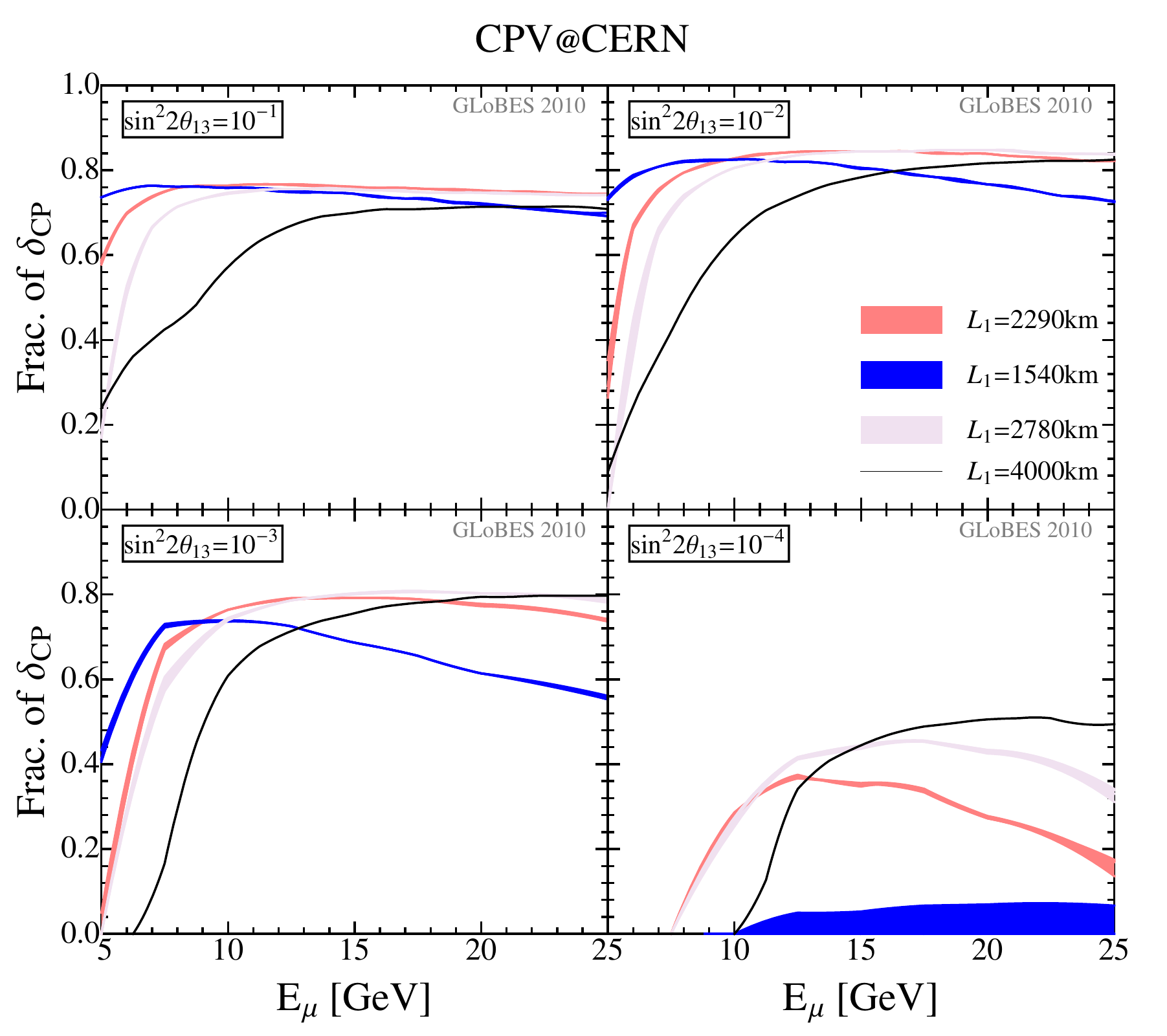}
 \mycaption{CPV discovery reach at CERN as a function of $E_\mu$ for all CERN-options in \Tab~\ref{tab:combine} at the $3 \sigma$ CL for different values of (true) $\stheta$ (as given in the panels).  Here we assume SF=1.0 and two 50 kt detectors. We group the different baseline combinations according to the shorter baseline $L_1$, as shown in the legend. The bands basically reflect the variation of the second baseline. The IDS-NF baseline combination is shown by the black curves with $L_1=4000$ km and $L_2=7500$~km.}
 \label{fig:CERN}
\end{figure}

\begin{figure}[tp]
 \centering
 \includegraphics[width=0.8\textwidth]{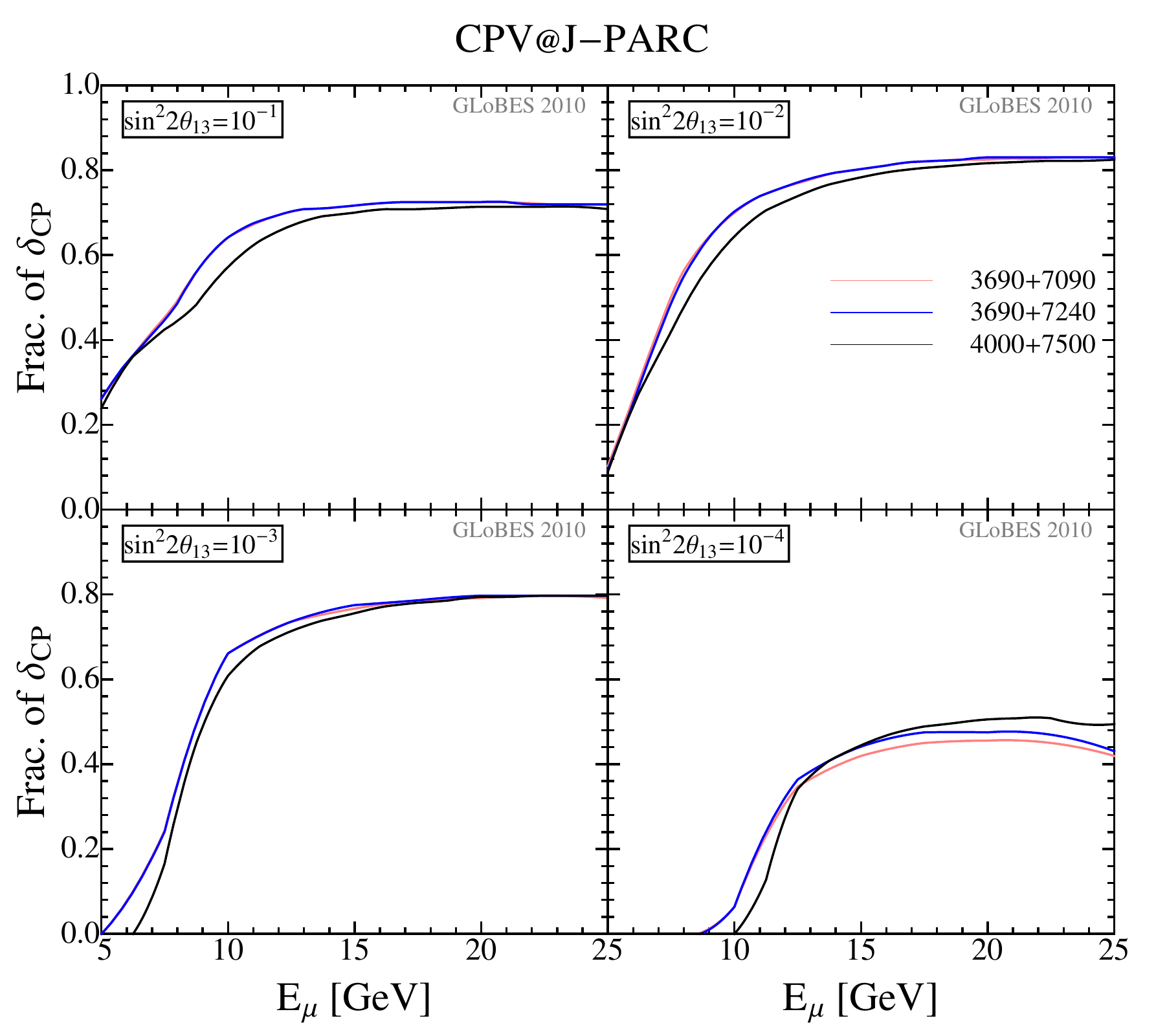}
 \mycaption{CPV discovery reach at J-PARC as a function of $E_\mu$ for all J-PARC-options in \Tab~\ref{tab:combine} at the $3 \sigma$ CL for different values of (true) $\stheta$ (as given in the panels).  Here we assume SF=1.0 and two 50 kt detectors. The IDS-NF baseline combination is shown by the black curves with $L_1=4000$ km and $L_2=7500$~km.}
 \label{fig:JPARC}
\end{figure}

\begin{figure}[tp]
 \centering
 \includegraphics[width=\textwidth]{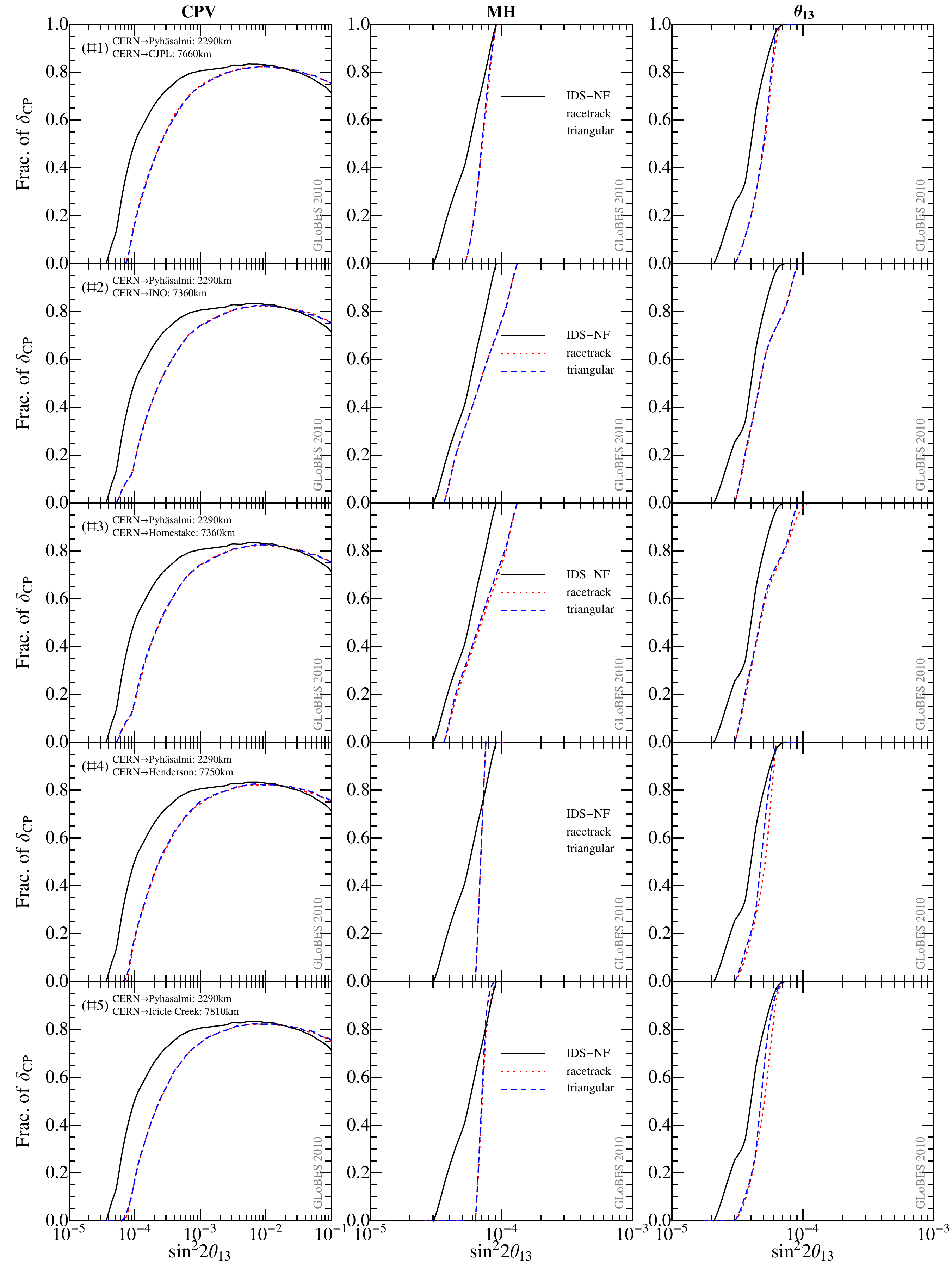}
 \mycaption{Discovery reach for CPV, MH, and $\theta_{13}$ for the combinations listed in \Tab~\ref{tab:combine}. Dotted (red) curves: SF=1 (racetrack-shaped storage rings), dashed (blue) curves: SF from \Tab~\ref{tab:combine} (triangular ring), black curves: IDS-NF baseline combination. Two
50 kt detectors used, $3 \sigma$ CL.}
 \label{fig:sample-1}
\end{figure}
\begin{figure}[tp]
 \centering
 \includegraphics[width=\textwidth]{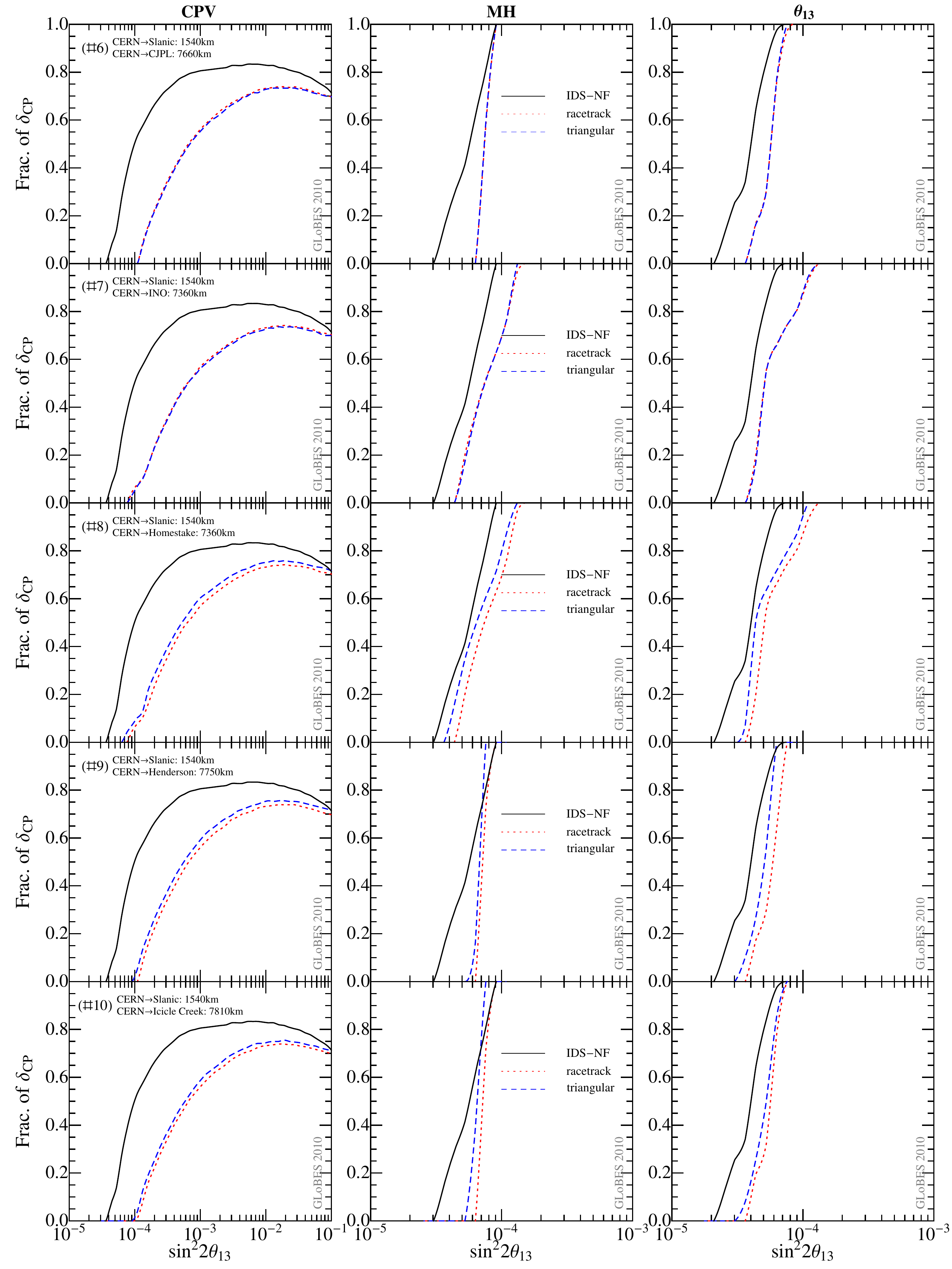}
 \label{fig:sample-2}
\end{figure}

\begin{figure}[tp]
 \centering
 \includegraphics[width=\textwidth]{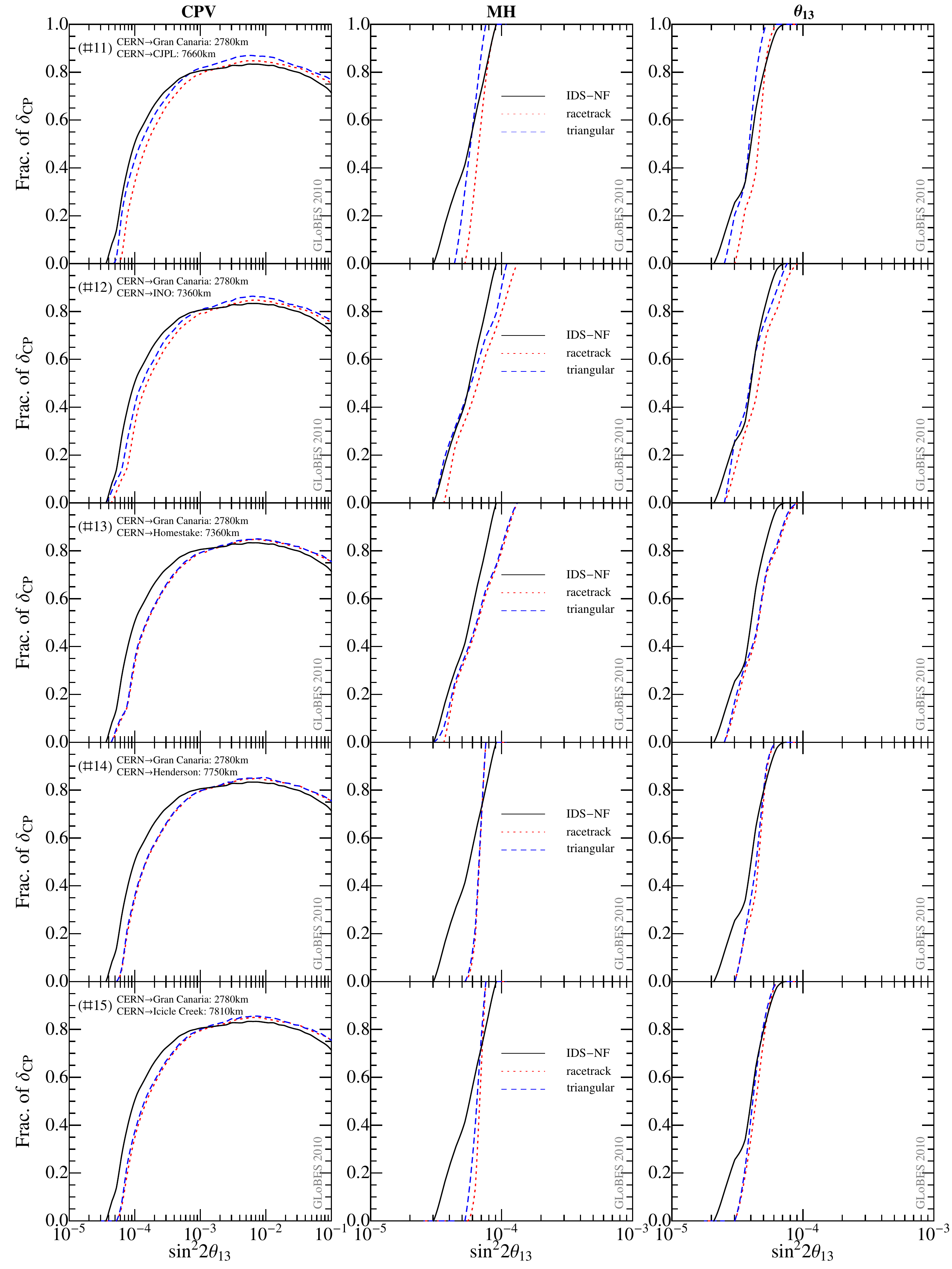}
 \label{fig:sample-3}
\end{figure}

\begin{figure}[tp]
 \centering
 \includegraphics[width=\textwidth]{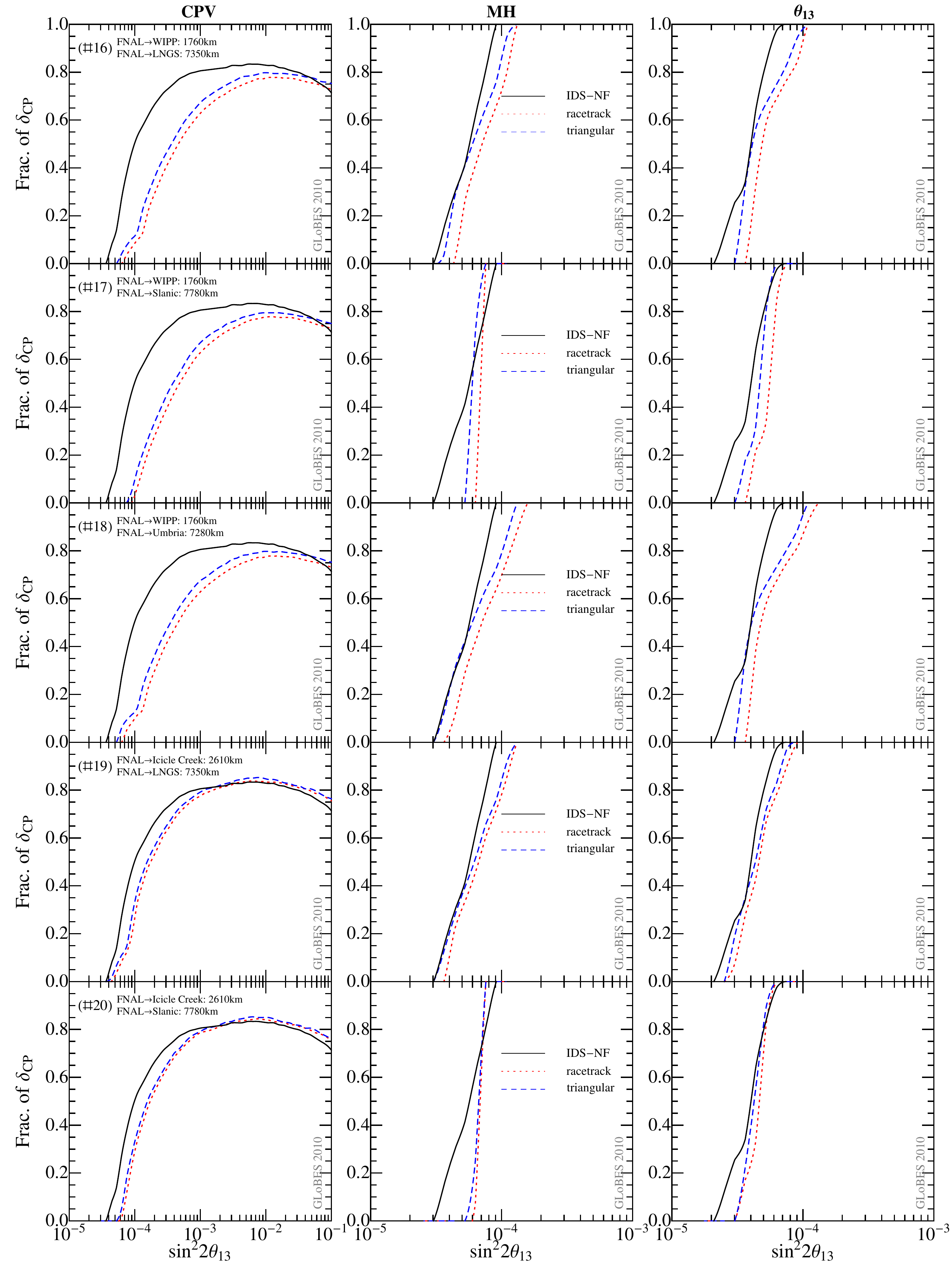}
 \label{fig:sample-4}
\end{figure}

\begin{figure}[tp]
 \centering
 \includegraphics[width=\textwidth]{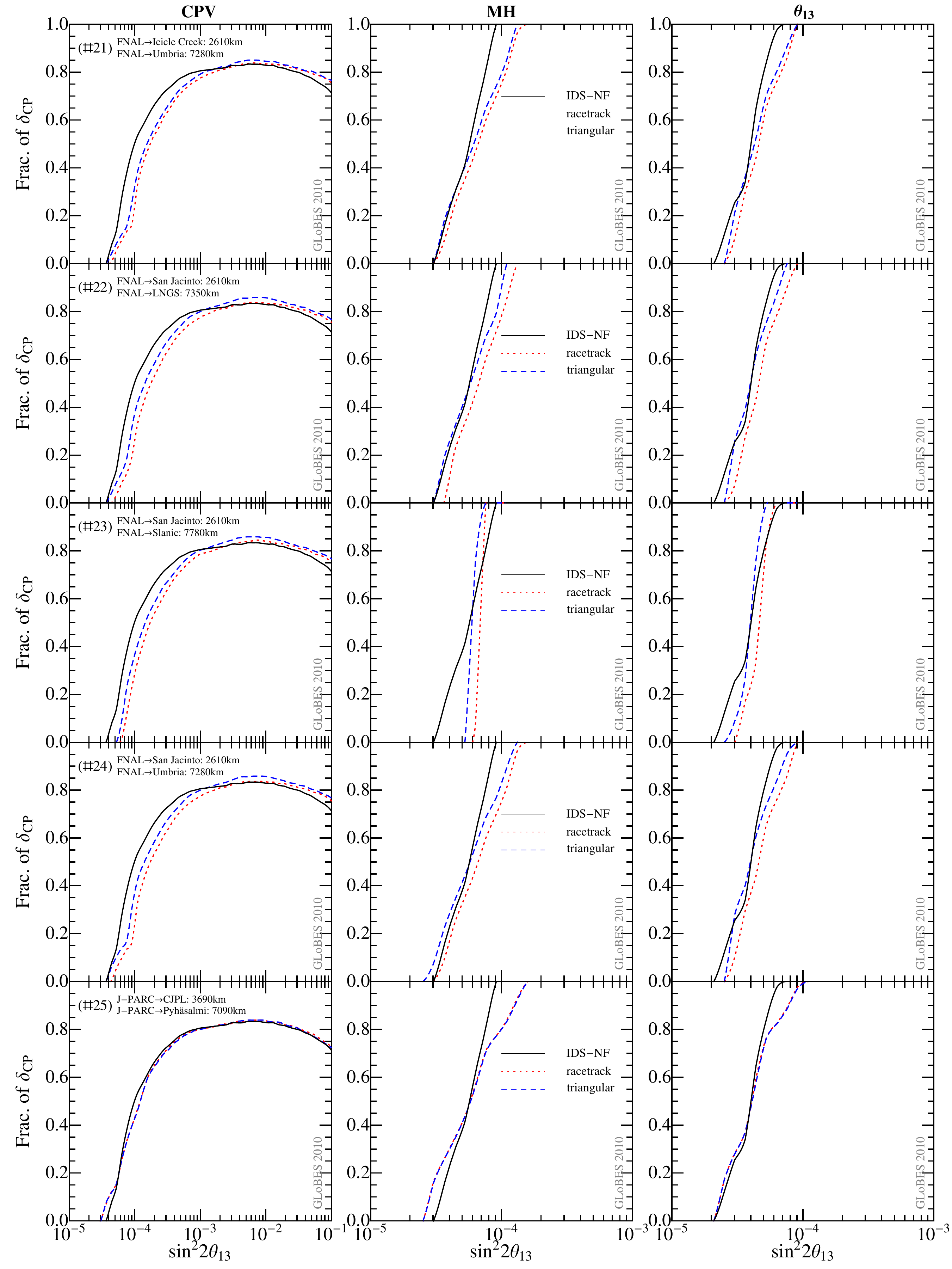}
 \label{fig:sample-5}
\end{figure}

\begin{figure}[tp]
 \centering
 \includegraphics[width=\textwidth]{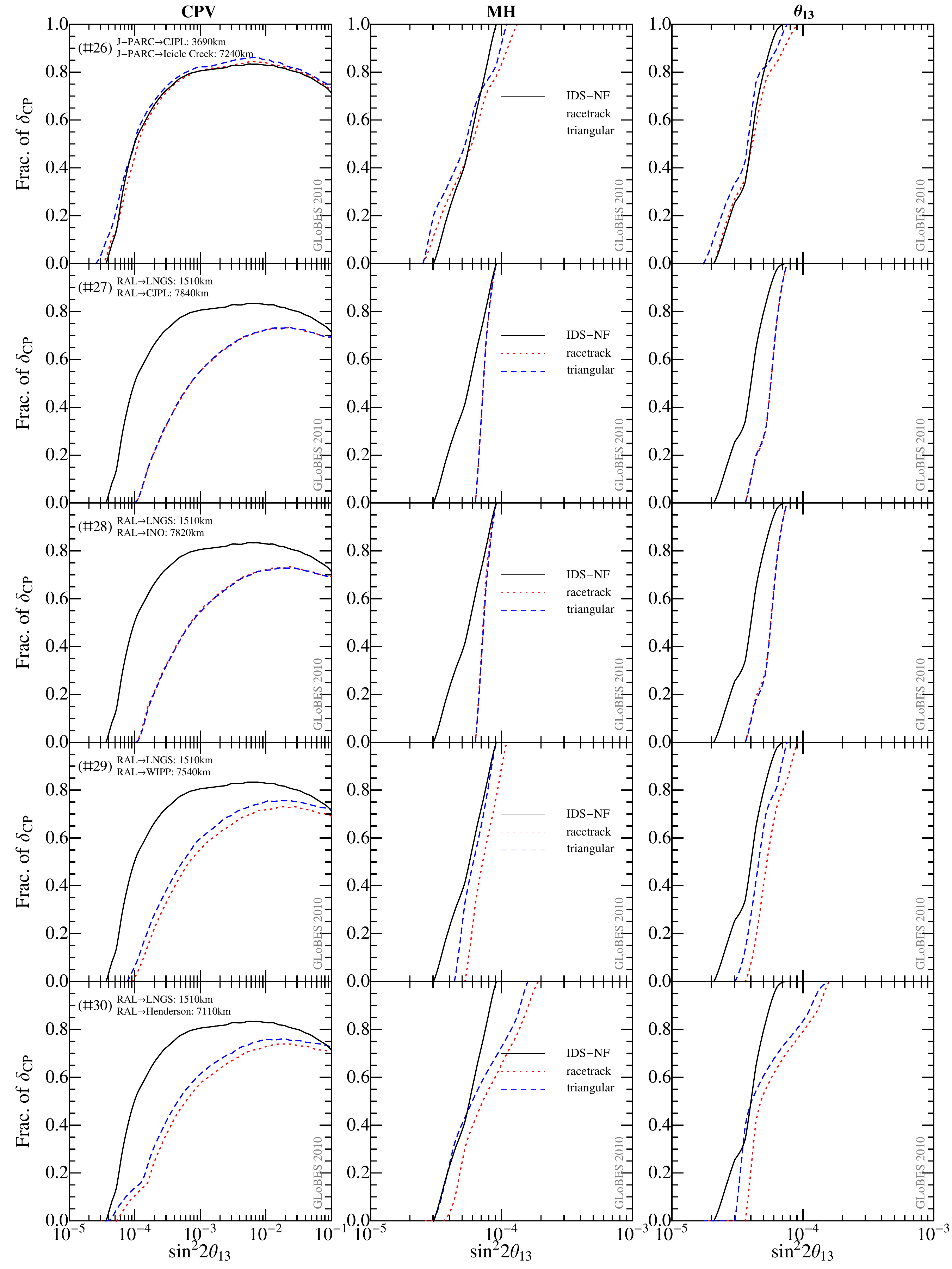}
 \label{fig:sample-6}
\end{figure}

\begin{figure}[tp]
 \centering
 \includegraphics[width=\textwidth]{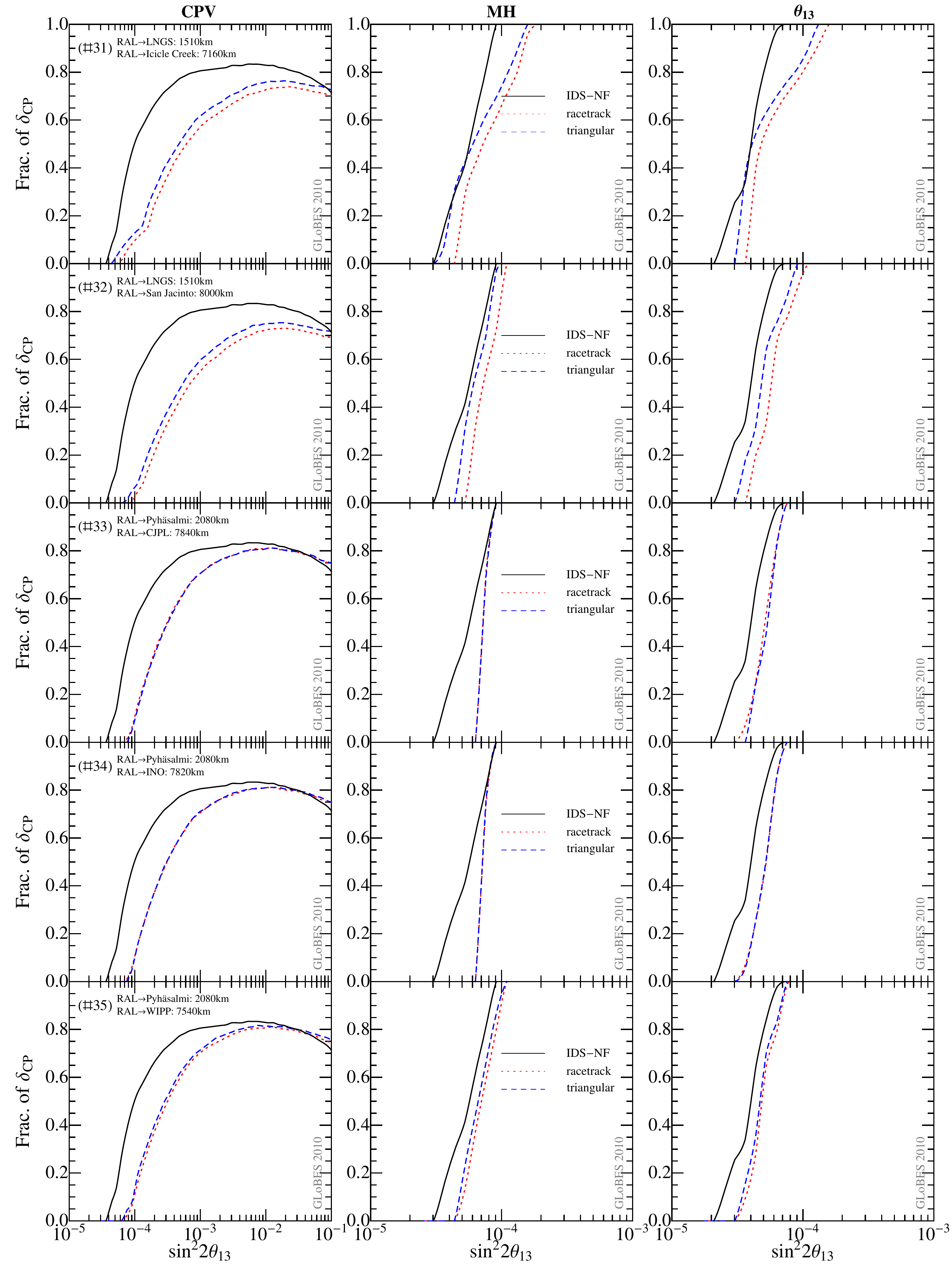}
 \label{fig:sample-7}
\end{figure}

\begin{figure}[tp]
 \centering
 \includegraphics[width=\textwidth]{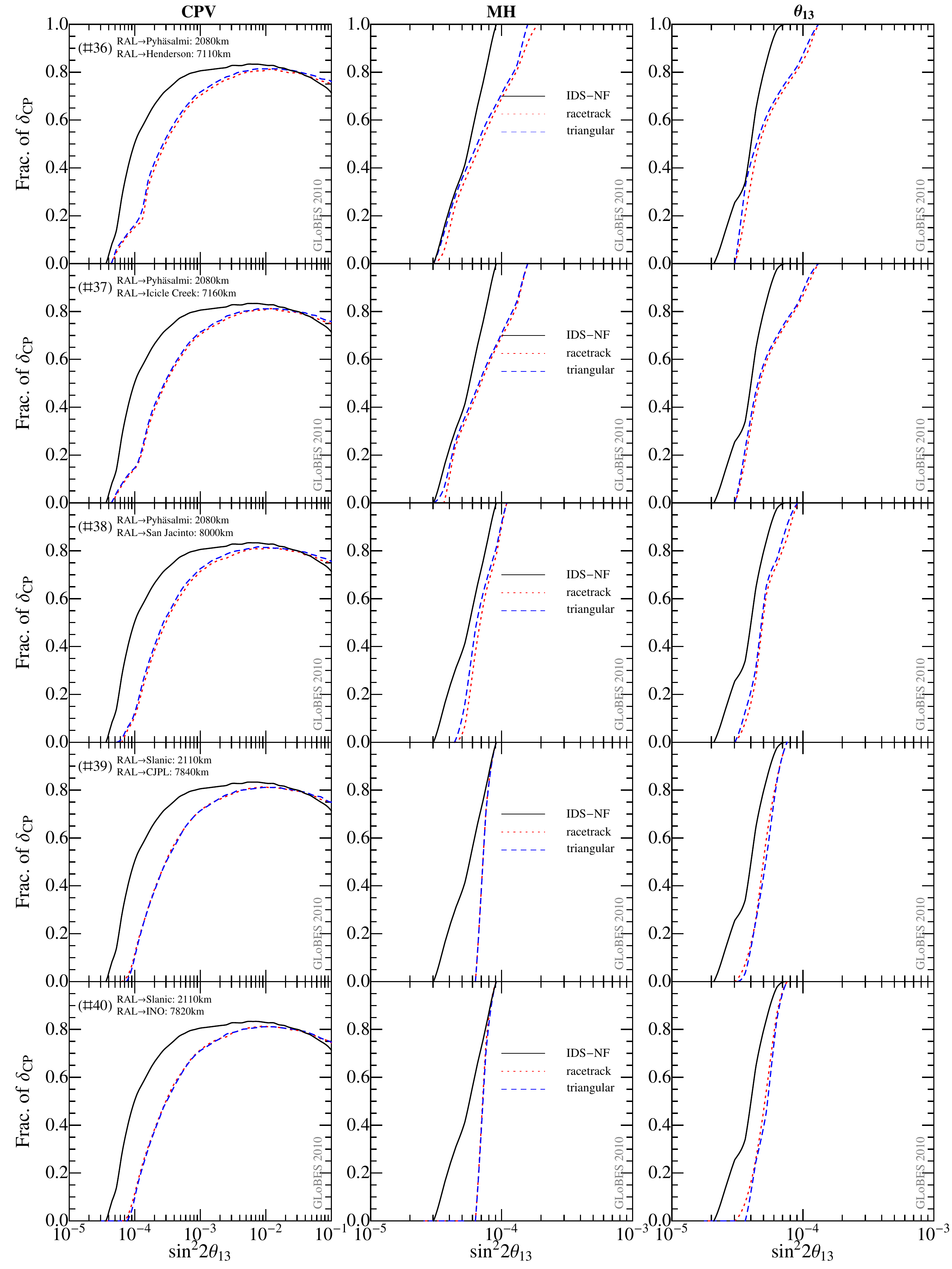}
 \label{fig:sample-8}
\end{figure}

\begin{figure}[tp]
 \centering
 \includegraphics[width=\textwidth]{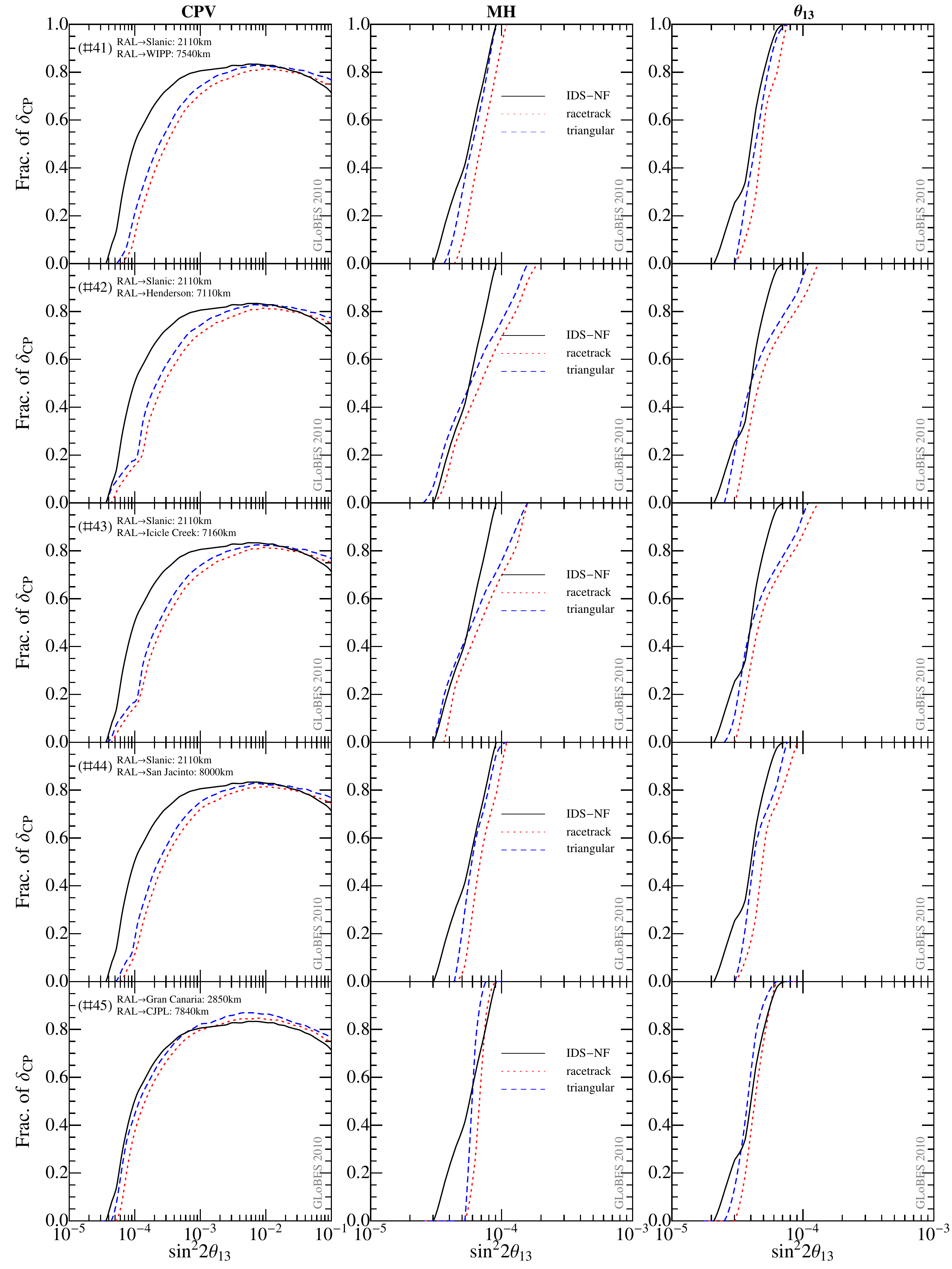}
 \label{fig:sample-9}
\end{figure}

\begin{figure}[tp]
 \centering
 \includegraphics[width=\textwidth]{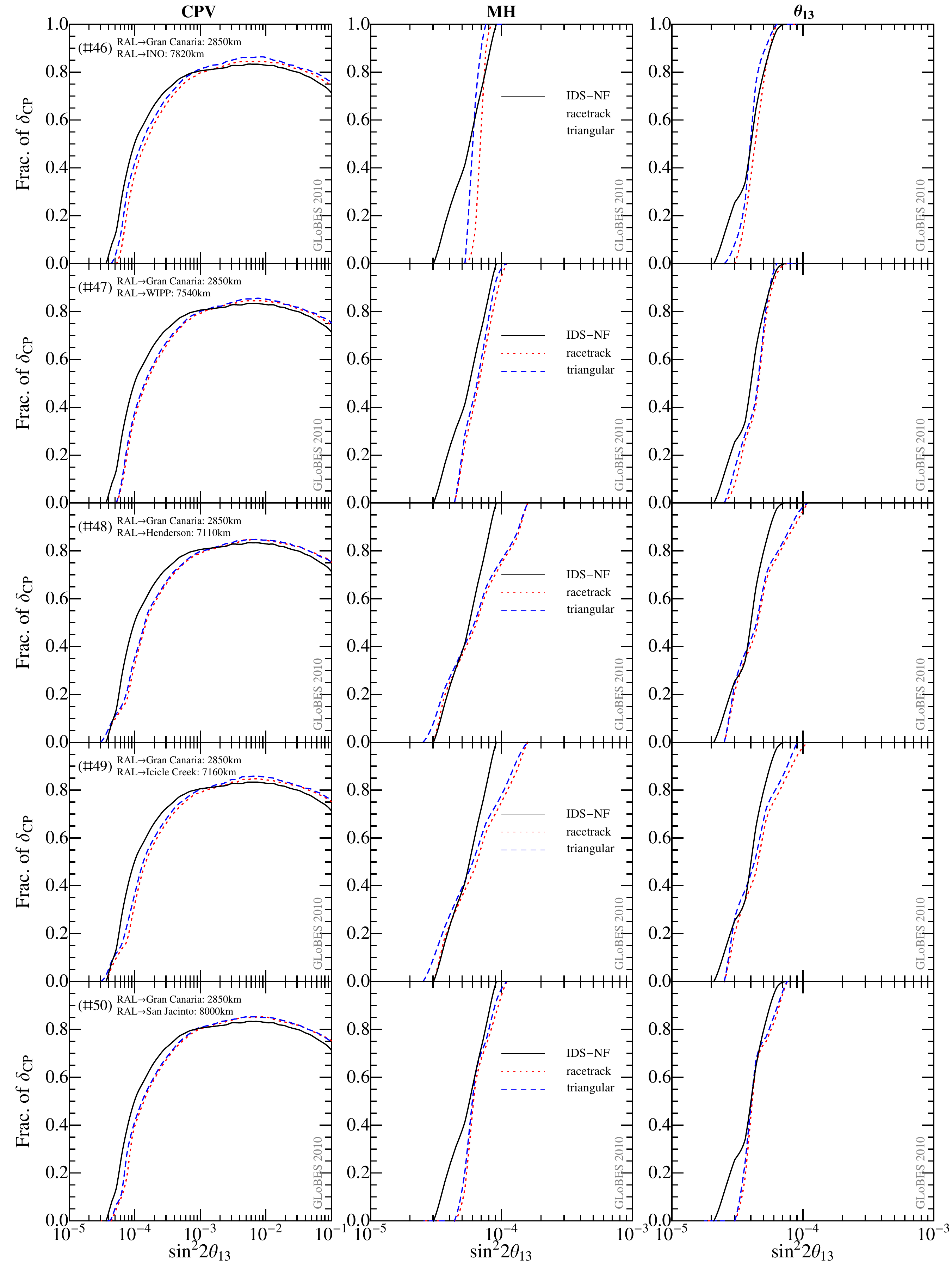}
 \label{fig:sample-10}
\end{figure}

Here we first of all show the figures similar to \figu{RAL} for the different accelerator laboratories for the sake of completeness: \figu{FNAL}, \figu{CERN}, and \figu{JPARC}. Although there are quantitative differences, there are no qualitatively new insights, apart from \figu{FNAL}. Here the FNAL-Homestake option is shown separately, which indeed peaks at even lower $E_\mu \simeq 5 \, \mathrm{GeV}$ for large $\stheta$.

In \figu{sample-1} (see also following pages), we show the performances of the discussed combinations (\#1 to \#50) from \Tab~\ref{tab:combine}. In each case, we compare the performances with racetracks (SF=1, dotted curves), triangular-shaped geometry (SF from \Tab~\ref{tab:combine}, dashed curves), and IDS combination (SF=1, solid curves). The main results of these figures are already discussed in \Sec~\ref{sec:siteperf}. The data points for the individual curves can be obtained at \Ref~\cite{dpage}.

\section{Note on the ``bi-magic'' baseline}
\label{app:bimagic}

\begin{figure}[t]
\begin{center}
\includegraphics[width=\columnwidth]{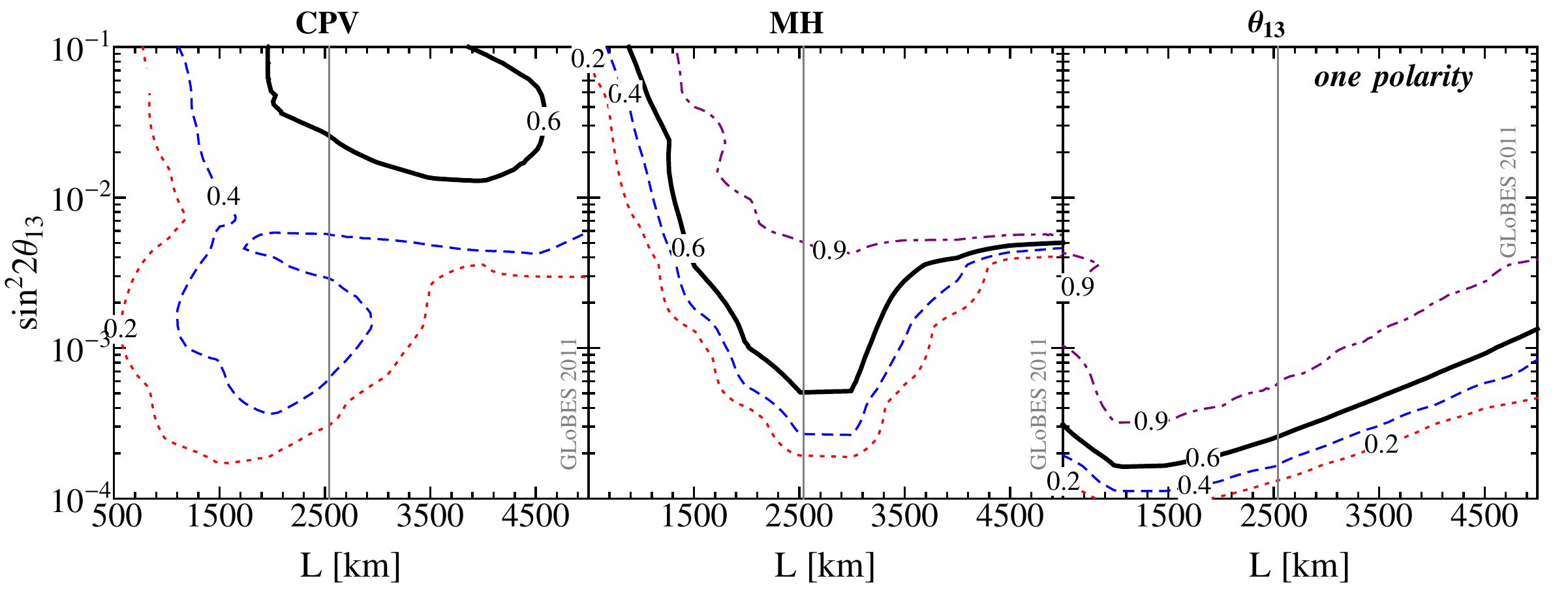}
\includegraphics[width=\columnwidth]{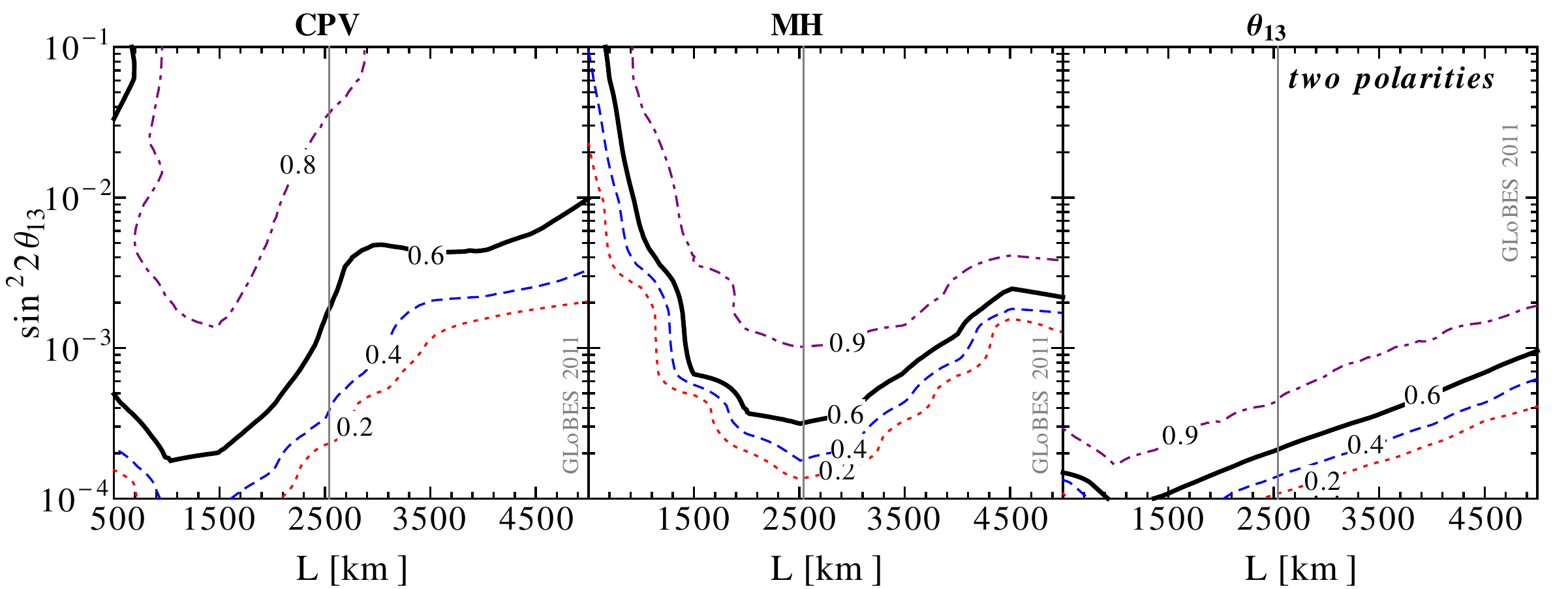}
\end{center}
\caption{\label{fig:master} Discovery reach in $\stheta$ as a function of baseline for CPV, MH, and $\theta_{13}$ discovery and specific fractions of $\deltacp$ (contours) for a TASD, see \Ref~\cite{Dighe:2010js} for details. The upper row is for one polarity ($\mu^+$ stored), the lower for two polarities. The results are shown at $3 \sigma$ CL with true normal hierarchy.} 
\end{figure}

In \Ref~\cite{Dighe:2010js}, the effect that the dependence on $\deltacp$ at a particular baseline and energy disappears (``bi-magic baseline'', $L \simeq 2540 \, \mathrm{km}$) for a chosen mass hierarchy, has been studied for a low energy neutrino factory together with a magnetized  totally active scintillator detector (TASD), which has somewhat different characteristics than the detector in this study (somewhat better efficiencies at low energies and a better energy resolution).  See also \Ref~\cite{Raut:2009jj} for the bi-magic baseline in the context of a superbeam, and \Ref~\cite{Barger:2006vy} for its baseline optimization.
We have reproduced the simulation in \Ref~\cite{Dighe:2010js} using the same beam and detector parameters, and we have studied the baseline dependence, see, \figu{master}, for the CPV, MH, and $\theta_{13}$ discovery reaches.
In this figure, specific values for the fraction of $\deltacp$ (contours) have been chosen to show the effect of the bi-magic baseline.

Our main observations from \figu{master} with respect to \Ref~\cite{Dighe:2010js} can be summarized as follows:
\begin{enumerate}
\item
 The optimal baseline choice depends on the performance indicator, the bi-magic baseline is mostly preferred for the MH.  Measuring CPV would clearly lead to a different baseline optimization.
\item
 There is clear preference for using both muon polarities, which will lead to a much better absolute sensitivity. 
\item
 There is no particular preference for exactly this baseline value, for none of the performance indicators, in the sense of the ``magic baseline''~\cite{Huber:2003ak} where a clear, narrow dip can be seen in the numerical study (see Figs.~3 and~6 in \Ref~\cite{Gandhi:2006gu}).
In addition, the ``bi-magic effect'' cannot be clearly attributed to the particular suggested energy windows. For example, if one masks the bins around the bi-magic energies (0.3 GeV windows around 1.9 and 3.3~GeV), the sensitivity is hardly affected.
\end{enumerate}
We have tested that our observations do not rely on the true hierarchy, or the energy resolution of the detector. 

In conclusion, the baseline of 2540~km is a good choice for the TASD if one wants to measure the MH as primary performance indicator for as small as possible values of $\stheta$. There is, however, no preference of this exact baseline value, instead a relatively wide baseline window is allowed. In addition, if CPV is considered as most important, a different baseline optimization will be clearly preferred.  

Finally, note that the logic of our paper is different: we have re-established the one baseline option of the low energy Neutrino Factory  for $\stheta \gtrsim 0.01$. In this case, the mass hierarchy can be measured for all values of $\deltacp$ for $L \gtrsim 1 \, 500 \, \mathrm{km}$ for the considered TASD (in the most pessimistic case $\stheta=0.01$). This means that the baseline optimization will be driven by the CPV measurement under the boundary condition of a minimal baseline for the mass hierarchy, which is illustrated in Figs.~1 and~3 of \Ref~\cite{Tang:2009wp}. From \figu{master} (lower left panel), we can read off that for $\stheta=0.01$, the smallest considered value for the LENF, the optimal baseline for both MH and CPV is then $1 \, 500 \, \mathrm{km} \lesssim L \lesssim 2 \, 100 \, \mathrm{km}$ (requiring a fraction of $\deltacp$ of 80\% for CPV and 100\% for MH), which is roughly consistent to the MIND detector, see \figu{LvsE} upper right panel. On the other hand, for $\stheta \lesssim 0.01$ ($\theta_{13}$ not discovered by the next generation of experiments), a two baseline HENF will be the optimal choice, and the MH discovery will be driven by the very long baseline.

\end{appendix}

\clearpage
\bibliographystyle{apsrev}
\bibliography{references}

\end{document}